\documentclass[longauth]{aa}  
\usepackage{subfigure}
\usepackage{graphicx}
\usepackage{hyperref}
\usepackage{comment}
\usepackage{multirow}
\usepackage{caption}

\usepackage{makecell}

\usepackage{tabularx}

\usepackage{txfonts}
\begin{document}


\title{Galaxy Morphology from $z\sim6$ through the eyes of JWST}

   \author{M. Huertas-Company,
          \inst{1,2,3,4}
          K. G. Iyer,
          \inst{5}
          E. Angeloudi,
          \inst{1,4}
          M. B. Bagley,
          \inst{7}
          S. L. Finkelstein,
          \inst{7}
          J. Kartaltepe,
          \inst{18}
          R. Sarmiento,
          \inst{1,4}
          J. Vega-Ferrero,
          \inst{1,4,9,10}
          P. Arrabal Haro,
          \inst{6}
          P. Behroozi,
          \inst{8,9}
          F. Buitrago,
          \inst{10,11}
          Y. Cheng,
          \inst{12}
          L. Costantin,
          \inst{13}
          A. Dekel,
          \inst{14,15}
          M. Dickinson,
          \inst{6}
          D. Elbaz,
          \inst{16}  
          N. A. Grogin,
          \inst{19}
          N. P. Hathi,
          \inst{19}
          B. W. Holwerda,
          \inst{17}
          A. M. Koekemoer,
          \inst{19}
          R. A. Lucas,
          \inst{19}
          C. Papovich,
          \inst{20,21}
          P. G. P\'erez-Gonz\'alez,
          \inst{13}
          N. Pirzkal,
          \inst{22}       
          L-M. Seill\'e,
          \inst{23}
          A. de la Vega,
          \inst{24}
          S. Wuyts,
          \inst{25}
          G. Yang,
          \inst{26,27}
          L. Y. A. Yung,
          \inst{28}          
          }

   \institute{Instituto de Astrof\'isica de Canarias (IAC), La Laguna, E-38205, Spain\\
              \email{mhuertas@iac.es}
         \and
             Observatoire de Paris, LERMA, PSL University, 61 avenue de l'Observatoire, F-75014 Paris, France
          \and  
          Universit\'e Paris-Cit\'e, 5 Rue Thomas Mann, 75014 Paris, France
          \and
          Universidad de La Laguna. Avda. Astrof\'{i}sico Fco. Sanchez, La Laguna, Tenerife, Spain
            \and
          Hubble Fellow, Columbia Astrophysics Laboratory, Columbia University, 550 West 120th Street, New York, NY 10027, USA
          \and 
          NSF's National Optical-Infrared Astronomy Research Laboratory, 950 N. Cherry Ave., Tucson, AZ 85719, USA
          \and
          Department of Astronomy, The University of Texas at Austin, Austin, TX, USA
          \and
          Department of Astronomy and Steward Observatory, University of Arizona, Tucson, AZ 85721, USA
          \and
          Division of Science, National Astronomical Observatory of Japan, 2-21-1 Osawa, Mitaka, Tokyo 181-8588, Japan
          \and 
          Departamento de F\'{i}sica Te\'{o}rica, At\'{o}mica y \'{O}ptica, Universidad de Valladolid, 47011 Valladolid, Spain
          \and 
          Instituto de Astrof\'{i}sica e Ci\^{e}ncias do Espa\c{c}o, Universidade de Lisboa, OAL, Tapada da Ajuda, PT1349-018 Lisbon, Portugal
          \and
          University of Massachusetts Amherst, 710 North Pleasant Street, Amherst, MA 01003-9305, USA
          \and 
          Centro de Astrobiolog\'{\i}a (CAB), CSIC-INTA, Ctra. de Ajalvir km 4, Torrej\'on de Ardoz, E-28850, Madrid, Spain
          \and
          Centre for Astrophysics and Planetary Science, Racah Institute of Physics, The Hebrew University, Jerusalem, 91904, Israel
          \and 
          Santa Cruz Institute for Particle Physics, University of California, Santa Cruz, CA 95064, USA
          \and 
          Laboratoire AIM-Paris-Saclay, CEA/DRF/Irfu - CNRS - Universit\'e Paris Cit\'e, CEA-Saclay, pt courrier 131, F-91191 Gif-surYvette, France
          \and 
          Physics \& Astronomy Department, University of Louisville, 40292 KY, Louisville, USA
          \and 
          Laboratory for Multiwavelength Astrophysics, School of Physics and Astronomy, Rochester Institute of Technology, 84 Lomb Memorial Drive, Rochester, NY 14623, USA
          \and
          Space Telescope Science Institute, 3700 San Martin Dr., Baltimore, MD 21218, USA
          \and
          Department of Physics and Astronomy, Texas A\&M University, College Station, TX, 77843-4242 USA
          \and 
          George P.\ and Cynthia Woods Mitchell Institute for Fundamental Physics and Astronomy, Texas A\&M University, College Station, TX, 77843-4242 USA
          \and 
          ESA/AURA Space Telescope Science Institute
          \and
          Aix Marseille Univ, CNRS, CNES, LAM Marseille, France
          \and 
          Department of Physics and Astronomy, University of California, 900 University Ave, Riverside, CA 92521, USA
          \and 
          Department of Physics, University of Bath, Claverton Down, Bath BA2 7AY, UK
          \and 
          Kapteyn Astronomical Institute, University of Groningen, P.O. Box 800, 9700 AV Groningen, The Netherlands
          \and
          SRON Netherlands Institute for Space Research, Postbus 800, 9700 AV Groningen, The Netherlands
          \and 
          Astrophysics Science Division, NASA Goddard Space Flight Center, 8800 Greenbelt Rd, Greenbelt, MD 20771, USA
             }

   \date{Received September 15, 1996; accepted March 16, 1997}

 \abstract
{The James Webb Space Telescope's unprecedented combination of sensitivity, spatial resolution, and infrared coverage enables a new exploration of galaxy morphology across most of cosmic history.}
{We analyze the Near Infrared ($\sim0.8-1\mu$m) rest-frame morphologies of galaxies with $\log M_*/M_\odot>9$ in the redshift range $0<z<6$, compare with previous HST-based results and release the first JWST-based morphological catalog of $\sim20,000$ galaxies in the CEERS survey.}
{Galaxies are classified into four main broad classes -- spheroid, disk+spheroid, disk, and disturbed -- based on imaging with four filters -- $F150W$, $F200W$, $F356W$, and $F444W$ -- using Convolutional Neural Networks trained on HST/WFC3 labeled images and domain-adapted to JWST/NIRCam.}
{We find that $\sim90\%$ and $\sim75\%$ of galaxies at $z<3$ have the same early/late and regular/irregular classification, respectively, in JWST and HST imaging when considering similar wavelengths.
For small (large) and faint objects, JWST-based classifications tend to systematically present less bulge-dominated systems (peculiar galaxies) than HST-based ones, but the impact on the reported evolution of morphological fractions is less than $\sim10\%$. Using JWST-based morphologies at the same rest-frame wavelength ($\sim0.8-1\mu$m), we confirm an increase in peculiar galaxies and a decrease in bulge-dominated galaxies with redshift, as reported in previous HST-based works, suggesting that the stellar mass distribution, in addition to light distribution, is more disturbed in the early universe. However, we find that undisturbed disk-like systems already dominate the high-mass end of the late-type galaxy population ($\log M_*/M_\odot>10.5$) at $z\sim5$, and bulge-dominated galaxies also exist at these early epochs, confirming a rich and evolved morphological diversity of galaxies $\sim1$ Gyr after the Big Bang. Finally, we find that the morphology-quenching relation is already in place for massive galaxies at $z>3$, with massive quiescent galaxies ($\log M_*/M_\odot>10.5$) being predominantly bulge-dominated. }
 {}

   \keywords{Astronomical Databases: Catalogs --
                Galaxies: evolution --
                Galaxies: high-redshift -- Galaxies: statistics -- Galaxies: structure
               }

   \titlerunning{CEERS morphologies}
\authorrunning{Huertas-Company et al.}
   \maketitle


\section{Introduction}

  Understanding how galaxy diversity emerges across cosmic time is one of the major goals of the field of galaxy formation. Questions such as how and when stellar disks form, what the main drivers of bulge growth are, and how and when galaxy morphology and star-formation got connected remain largely unanswered despite significant progress in recent years (e.g., \citealp{ 2014ARA&A..52..291C,2020ARA&A..58..661F}). Galaxy morphology remains a key first-order proxy of galaxy diversity since it is a fast and cost-effective way of identifying different types of galaxies in large samples at different cosmic times, requiring only imaging. Up to now, almost all studies involving galaxy structure beyond the local Universe make use of the Hubble Space Telescope (HST), which was the only facility delivering high enough spatial resolution to study the structure of small high-redshift galaxies \citep[e.g.,][]{1996MNRAS.279L..47A, 2015ApJS..221....8H}. HST imaging has revealed that galaxies at $z>1$ tend to become more irregular in their light distribution \citep[e.g.,][]{2000ApJ...529..886C}, even if observed in the rest-frame optical more sensitive to main-sequence stars \citep[e.g.,][]{2016MNRAS.462.4495H}. It has also revealed that star-forming and quiescent galaxies have different stellar structures at all cosmic epochs probed so far \citep[e.g.,][]{2011ApJ...742...96W, 2014ApJ...788...28V,2021A&A...646A.151P,2022MNRAS.513..256D} and many others. The recent launch of the James Webb Space Telescope (JWST) opens a new window on galaxy morphology, especially at early epochs, by providing unprecedented spatial resolution and depth combined with infrared coverage that allows probing the optical rest-frame emission in the first billion years of the Universe's history. A few early works analyzing the first set of JWST images have started exploring the morphological diversity up to $z\sim6$ using visual classifications or deep learning models trained on HST images \citep[e.g.,][]{2022ApJ...938L...2F, 2022arXiv221001110F, 2022arXiv221014713K, 2023ApJ...942L..42R}. The common conclusion of these initial studies is that there might be more disk-like galaxies at high redshift than previously inferred with HST, although the exact nature of these disk-like galaxies still needs to be confirmed \citep{2023arXiv230207277V}.

In this work, we take a step forward in understanding the morphological diversity of galaxies in the early Universe by providing the first and largest publicly available morphological classification of $\sim20,000$ galaxies selected in the JWST Cosmic Evolution Early Release Science (CEERS,~\citealp{2022ApJ...940L..55F,2023ApJ...946L..13F}) survey and observed in four different wavelengths ($F150W$, $F200W$, $F356W$, and $F444W$). Galaxies are classified into four broad morphological classes using a convolutional neural network (CNN) trained on HST-based classifications from the Cosmic Assembly Near-infrared Deep Extragalactic Legacy Survey (CANDELS, ~\citealp{2011ApJS..197...36K,2011ApJS..197...35G}) and domain-adapted to work on JWST images. We then use the new classifications to precisely quantify the changes in galaxy morphology when moving from HST to JWST, study the evolution of morphological fractions from $z\sim6$ to $z\sim0$ in the NIR rest-frame ($\sim0.8-1\mu m$) by using different filters at different redshifts, and revisit the morphology-quenching relation over $\sim90\%$ of the cosmic history.

  The paper proceeds as follows. In section~\ref{sec:data} we present the datasets used, namely the CEERS and the CANDELS surveys. Section~\ref{sec:method} describes the method used to estimate morphologies on JWST images and Section~\ref{sec:HST_JWST} systematically compares HST and JWST based galaxy morphologies. Section~\ref{sec:morph_evol} studies the evolution of morphological fractions between $z\sim0$ and $z\sim6$ as well as the morphology-quenching relation and Section~\ref{sec:disc} discusses some of the implications of the results presented. A summary is presented in Section~\ref{sec:summary}. Throughout this paper we use a Planck 2013 cosmology~\citep{2014A&A...571A..16P}.

\section{Data}
\label{sec:data}

\subsection{CEERS}
CEERS~\citep{2022ApJ...940L..55F,2023ApJ...946L..13F} is an Early Release Science (ERS) program (Proposal ID 1345, PI: Finkelstein) that has been observing the EGS (Extended Groth Strip,~\citealp{2007ApJ...660L...1D}) extragalactic deep field (one of the five CANDELS fields;~\citealp{2011ApJS..197...36K,2011ApJS..197...35G}) since June 2022, with data made available to the public immediately. For this work, we combined 10 NIRCam pointings observed in June and December 2022 in four different filters: F150W, F200W, F356W, and F444W. For the June pointings, we used the images from the latest CEERS public data release\footnote{\url{https://ceers.github.io}}. A detailed description of the reduction process can be found in~\cite{2022arXiv221102495B}. For the December data, we used internal data products from the CEERS team reduced with an analogous procedure.

In addition to imaging data, we also used photometric redshifts and physical properties of galaxies derived with Spectral Energy Distribution (SED) fitting. Photometric redshifts were derived using the EAZY code~\citep{2008ApJ...686.1503B}, as detailed in~\cite{2022ApJ...940L..55F}. The physical properties of galaxies - namely, stellar masses and star-formation rates for this work - have been determined using the \textsc{Dense Basis} method\footnote{\href{https://dense-basis.readthedocs.io/}{https://dense-basis.readthedocs.io/}} \citep{2019ApJ...879..116I}, an SED fitting method using a procedure similar to that in \citealt{2019ApJ...879..116I, 2021ApJ...913...45O, 2022ApJ...940L..55F, 2022ApJ...937L..35M, 2022arXiv221207540A}. The code performs a fully Bayesian inference of the Star Formaton History (SFH), dust attenuation, and chemical enrichment for each galaxy, using a fully non-parametric Gaussian process-based description for the SFH described in \citealt{2019ApJ...879..116I}. The model uses a Chabrier IMF \citep{2003PASP..115..763C}, a Calzetti dust attenuation law \citep{2000ApJ...533..682C}, and Madau IGM absorption \citep{1995ApJ...441...18M}. In the fitting process, a redshift prior taken from the confidence interval of the EAZY estimate is used. Setting explicit priors in SFH space using non-parametric SFHs has been shown to be robust against outshining due to younger stellar populations, which can otherwise bias estimates of masses and star formation rates \citep{2017ApJ...838..127I,2019ApJ...876....3L,2020ApJ...904...33L}.

.

\subsection{CANDELS}

We use publicly available images and data products from the CANDELS survey~\citep{2011ApJS..197...36K,2011ApJS..197...35G} from the five different fields~\citep{2013ApJS..206...10G,2013ApJS..207...24G,2017ApJS..228....7N,2017ApJS..229...32S,2019ApJS..243...22B}. For training the neural networks (see section~\ref{sec:method}), we use the morphological catalog of~\cite{2015ApJS..221....8H}, which provides neural network based morphological labels trained on the~\cite{2015ApJS..221...11K} visual classifications for all galaxies with $F160W<24.5$. We decided to use as ground truth, the labels from the CNN based classification of~\cite{2015ApJS..221....8H} instead of directly using the visual classifications from~\cite{2015ApJS..221...11K} because the former provides an homogeneous and larger sample. Given the excellent agreement shown in~\cite{2015ApJS..221....8H} we do not expect this choice to introduce any significant bias. For this work, we define 4 main morphological classes using the criteria also described in~\cite{2015ApJS..221....8H}, which have been demonstrated to provide reasonably clean classes. Namely:

\begin{itemize}
\item {\bf Class 1:} \emph{spheroids / pure bulges:}~~$f_{sph}>2/3$ AND $f_{disk}<2/3$ AND $f_{irr}<1/10$
\item {\bf Class 2:} \emph{disks:}~~$f_{sph}<2/3$ AND $f_{disk}>2/3$ AND $f_{irr}<1/10$
\item {\bf Class 3:} \emph{bulge+disk:}~~$f_{sph}>2/3$ AND $f_{disk}>2/3$ AND $f_{irr}<1/10$
\item {\bf Class 4:} \emph{irregulars / disturbed / peculiar:}~~ $f_{irr}>1/10$

\end{itemize}

where $f_{sph}$, $f_{disk}$ and $f_{irr}$ are estimates of the vote fractions of different classifiers for spheroid, disk and irregular features respectively (see~\citealp{2015ApJS..221...11K} and~\citealp{2015ApJS..221....8H} for more details). For the remainder of this work, we consider the labels corresponding to the four classes as ground truth. We will use the terms spheroids or pure bulges interchangeably to refer to the first class, and irregulars, disturbed, or peculiar to refer to the last class. Additionally, we will refer to the combination of classes 1 and 3 as early-type galaxies, and the combination of classes 2 and 4 as late-type galaxies. The inclusion of bulge+disks systems into the early-type class is motivated by the results of ~\cite{2015ApJS..221....8H} who showed that the bulge+disk class is mainly composed of bulge-dominated systems. We will discuss this further in the following sections. Because we have removed stars using one of the filters with highest spatial resolution, we do not include a point source classification, which would furthermore decrease the accuracy of our classification because of low statistics. 

It is important to note that the names given to the classes are solely used to identify different morphological properties based on images, and do not necessarily reflect the true physical nature of these objects. In particular, the irregular class can contain a variety of galaxies with different physical properties. The spheroid class generally refers to round and compact galaxies which do not necessarily imply kinematically hot systems, especially at high redshift.

\subsection{Sample Selection and Completeness}

\begin{figure}
 {\includegraphics[width=1\linewidth]{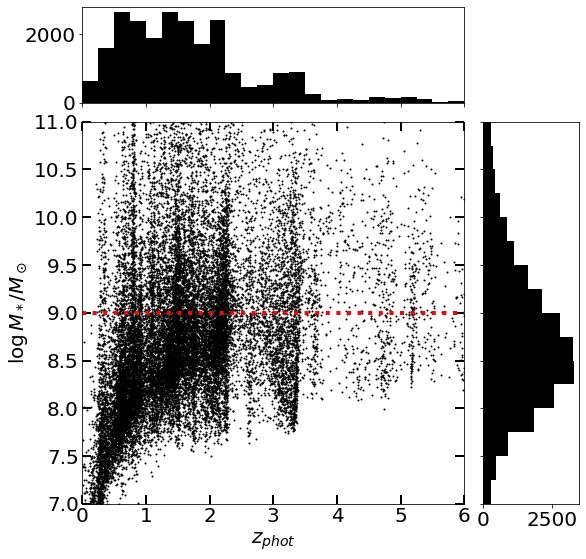}}
    \caption{Distribution of selected galaxies ($F200W<27$) in the stellar mass - photometric redshift plane. The dashed red line shows the $10^9$ solar mass limit used in this work for scientific analysis, whcih is well above the completeness limit.   } 
    \label{fig:m_z}
\end{figure}

We select all galaxies with $F200W[AB]<27$. The magnitude cut corresponds roughly to a Signal-to-Noise Ratio (SNR) which enables reliable morphological classification (see~\citealp{2022arXiv221014713K}). In addition to the apparent magnitude cut, we also remove obvious stars using a simple procedure described in Appendix~\ref{app:star-galaxy} using photometric parameters measured in the F200W filter. The final sample for which we estimate galaxy morphologies is then of $23,674$ galaxies. Figure~\ref{fig:m_z} shows the distribution of the selected galaxies in the stellar mass - photometric redshift plane. The lower envelope of the distribution can be used as a proxy for completeness which is estimated at roughly $10^{8.5} M_{\odot}$ over the considered redshift range. Although the released morphological catalog includes all galaxies, in the remaining of this work, we will primarily focus on galaxies more massive than $10^9$ solar masses - which is well above the completeness limit - so the main conclusions should not be affected by incompleteness.

\section{Inferring galaxy morphologies of CEERS galaxies}
\label{sec:method}

\subsection{Method}
Galaxy morphology is commonly estimated with supervised Convolutional Neural Networks (CNNs), which allow one to efficiently extract image features correlated with galaxy morphology (see the review by \citealp{2023PASA...40....1H}). The main bottleneck of such an approach is the training set, which needs to contain a large enough sample of annotated images, typically performed through visual inspection. However, this is a time-consuming process that potentially undermines the advantage of using machine learning. Therefore, several works have tried to find global features (e.g., \citealp{2022arXiv220611927W,2023arXiv230202005C}) or transfer a network trained on one dataset to another (e.g., \citealp{2019MNRAS.484...93D}).

Since there are very few available labels on JWST images, we use standard adversarial domain adaptation~\citep{2015arXiv150507818G} in this work to transfer existing labels on HST imaging from the CANDELS survey to JWST-CEERS. Adversarial domain adaptation has been successfully applied to astronomical data in the context of galaxy merger identification (e.g., \citealp{2020arXiv201103591C}).

The overall idea is that both labeled images from HST and unlabeled images from JWST are shown to the CNN. An adversarial term is added to the loss functions to push the features extracted from the labeled dataset to be close to the ones from the unlabeled one. This way, the network is prevented from learning specific features linked to a particular dataset. More precisely, in this work, we use the network architecture illustrated in Figure \ref{fig:cnn} and the following loss function:

\begin{figure}
   \includegraphics[width=1\linewidth]{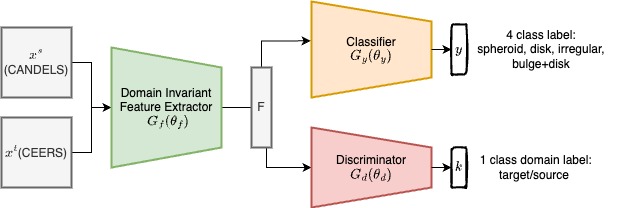}
    \caption{Schematic representation of the neural network architecture used for classifying CEERS galaxies in this work. A first CNN ($ G_f(\theta_f)$) is fed with both labeled and unlabeled CANDELS and CEERS stamps respectively. The computed features are then used as input for two additional CNNs: a discriminator ($G_d(\theta_d)$) which learns to distinguish stamps coming from the two datasets and a classifier ($G_y(\theta_y)$) which provides a classification in four main morphological classes. More details about the training strategy can be found in the text. } 
    \label{fig:cnn}
\end{figure}  

\begin{equation}
\begin{split}
    E(\theta_f,\theta_y,\theta_d) & =  \sum_{i=1..N} L_y(G_y(G_f(x_i^{s};\theta_f);\theta_y),y_i) \\
    & - \alpha L_d(G_d(G_f(x_i^t,x_i^s;\theta_f);\theta_d),k_i)
    \end {split}
    \label{eq:loss}
\end{equation}

where $G_f$, $G_y$, and $G_d$ are the feature extractor, classifier, and discriminator networks, respectively, with free parameters $\theta_f$, $\theta_y$, and $\theta_d$. $L_y$ and $L_d$ are the losses of the classifier and discriminator, which in our case are standard cross-entropy losses. $x^s$ and $x^t$ are source (CANDELS) and target (CEERS) images respectively. $y$ and $k$ are the morphological classes and the dataset class respectively. $\alpha$ is a scalar hyper-parameter that adjusts the weight between the two loss terms and hence acts as a trade-off between classification accuracy and domain invariance. In this work, we set $\alpha=1$ as in the original implementation. We investigated varying its value without significant impact on the final classification. Because there is a minus sign in equation \ref{eq:loss}, the network is effectively optimized so that the classification loss is minimized (i.e., classification accuracy) while the domain discriminator loss is maximized (i.e., domain invariance). Optimal values for the free parameters are thus given by:
\begin{equation}
\begin{split}
    (\hat{\theta_f},\hat{\theta_y}) & =\arg \min_{\theta_f,\theta_d} E(\theta_f,\theta_y,\hat{\theta_d}) \\
    \hat{\theta_d} & = \arg \max_{\theta_d} E(\hat{\theta_f},\hat{\theta_y},\theta_f)  
    \end {split}
\end{equation}

We train four identical networks using as source images, $\sim 50,000$ F160W stamps from the CANDELS survey distributed in five different fields with labels from~\cite{2015ApJS..221....8H} (see section~\ref{sec:data} for more details) and as target $F150W$, $F200W$, $F356W$ and $F444W$ until convergence ($\sim$ 50 epochs). The output of the classifier network is determined by a softmax layer which provides an array of size four corresponding to a measurement of the probability that a given galaxy image belongs to one of the four classes described in~\autoref{sec:data}. The stamp size is fixed to $32 \times 32$ pixels for both datasets - CEERS and CANDELS - which implies that the effective field of view is also different. We tried interpolating the CANDELS images as a preprocessing step to match the pixel scale of NIRCam, but it led to worse results. We thus decided to keep the different field of views and let the network learn the corrections.

The classification is ultimately done by applying the trained feature extractor ($G_f$) and classifier ($G_y$) to the CEERS target stamps ($x^t$). To account for uncertainty, we employ an ensemble of 10 separate trainings with different initial conditions and slightly different training sets. To estimate the final probability of a given morphological class, we take the average of the outputs from all 10 networks. The standard deviation is used to assess the robustness of the classification which we find to be typically below 0.1. Thus, unless otherwise noted, we define classes as the maximum of the average probability from the 10 classification networks. Excluding objects with uncertain classifications does not significantly impact the main findings. 

\subsection{Visual inspection}

\begin{figure*}
    \begin{minipage}[t]{0.5\linewidth}
        \centering
        \captionsetup{font=footnotesize}
        \caption*{Spheroids}
        \vspace{-10pt}
        \includegraphics[width=\linewidth]{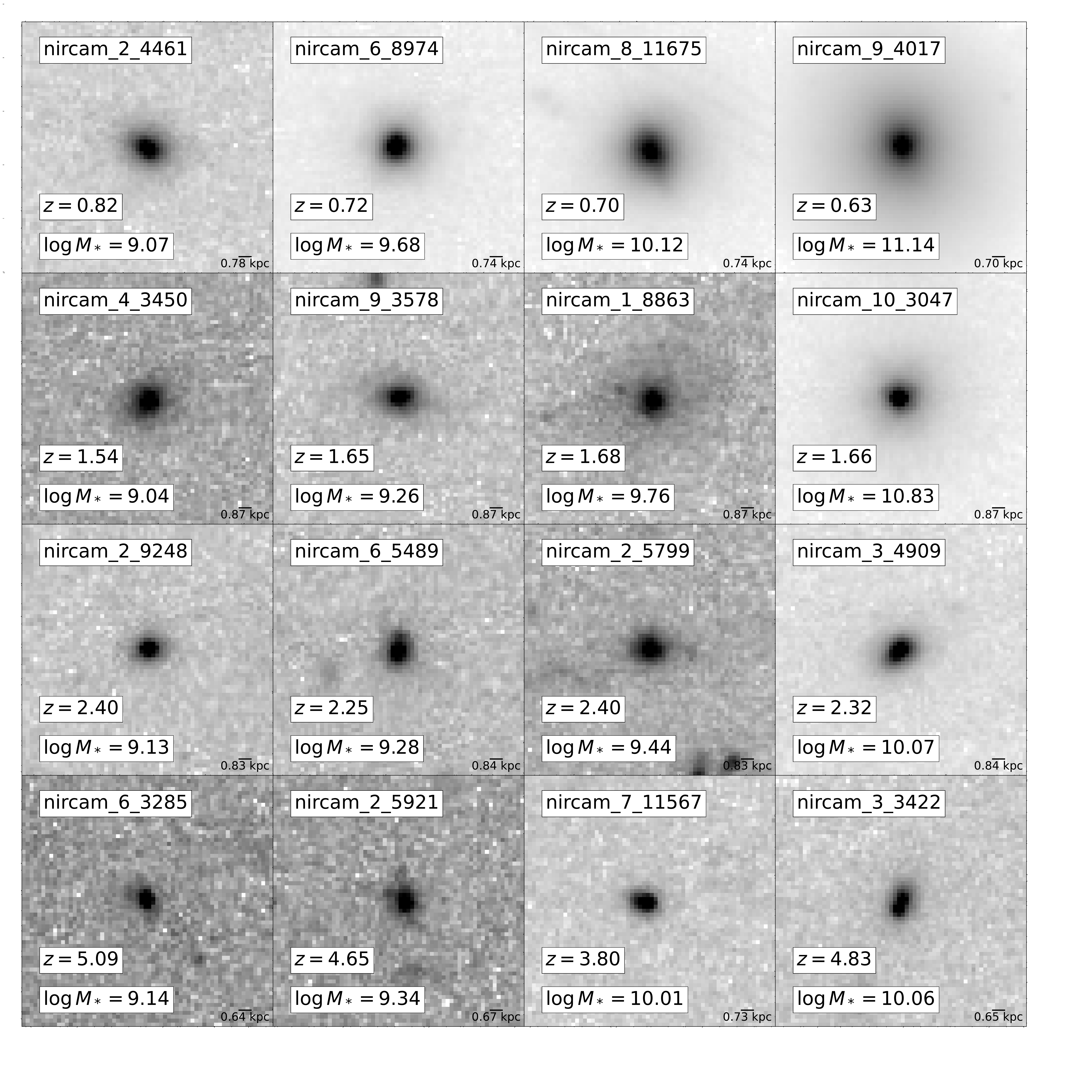}
    \end{minipage}%
    \begin{minipage}[t]{0.5\linewidth}
        \centering
        \captionsetup{font=footnotesize}
        \caption*{Disks}
        \vspace{-10pt}
        \includegraphics[width=\linewidth]{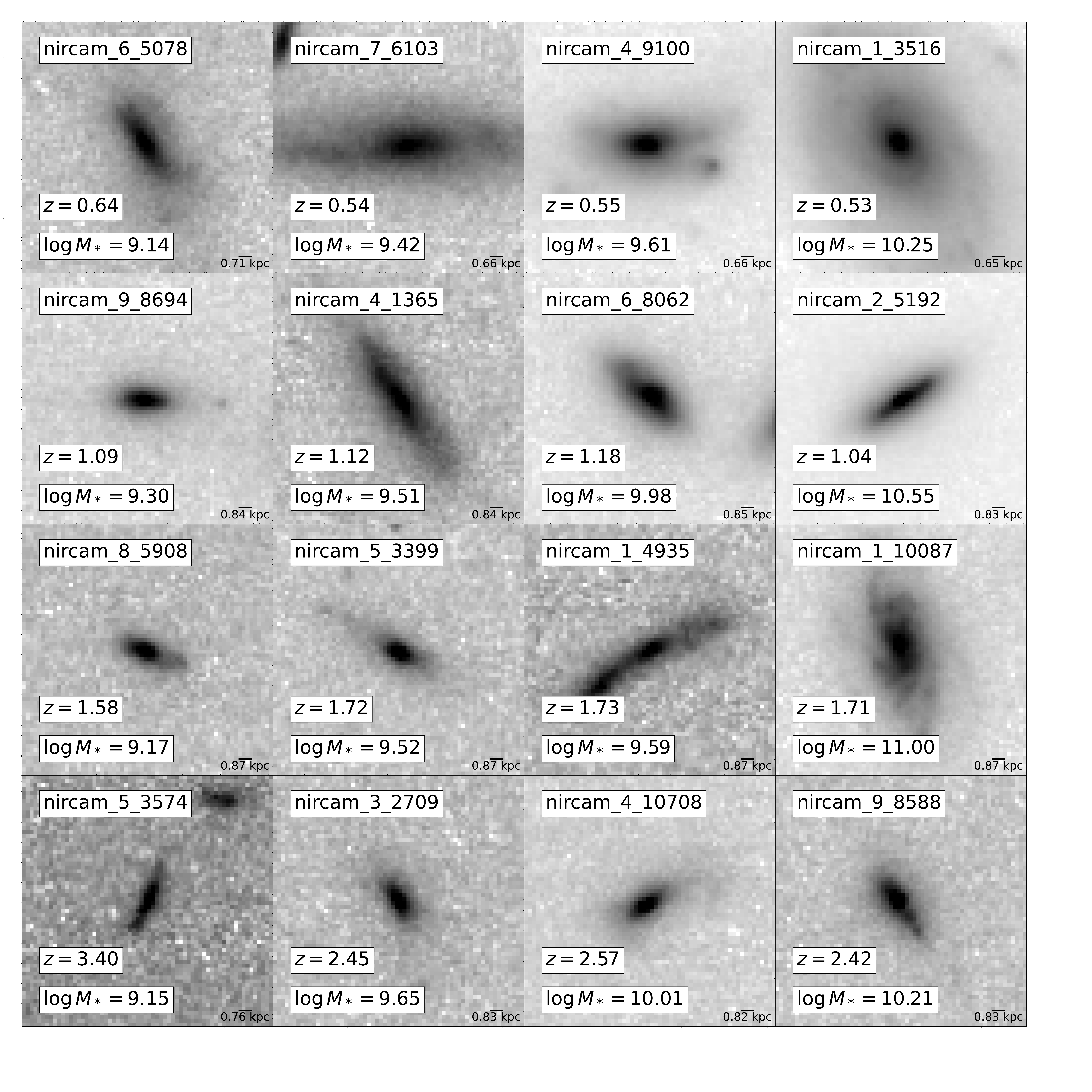}
    \end{minipage}
    
    \begin{minipage}[t]{0.5\linewidth}
        \centering
        \captionsetup{font=footnotesize}
        \caption*{Irregulars}
        \vspace{-10pt}
        \includegraphics[width=\linewidth]{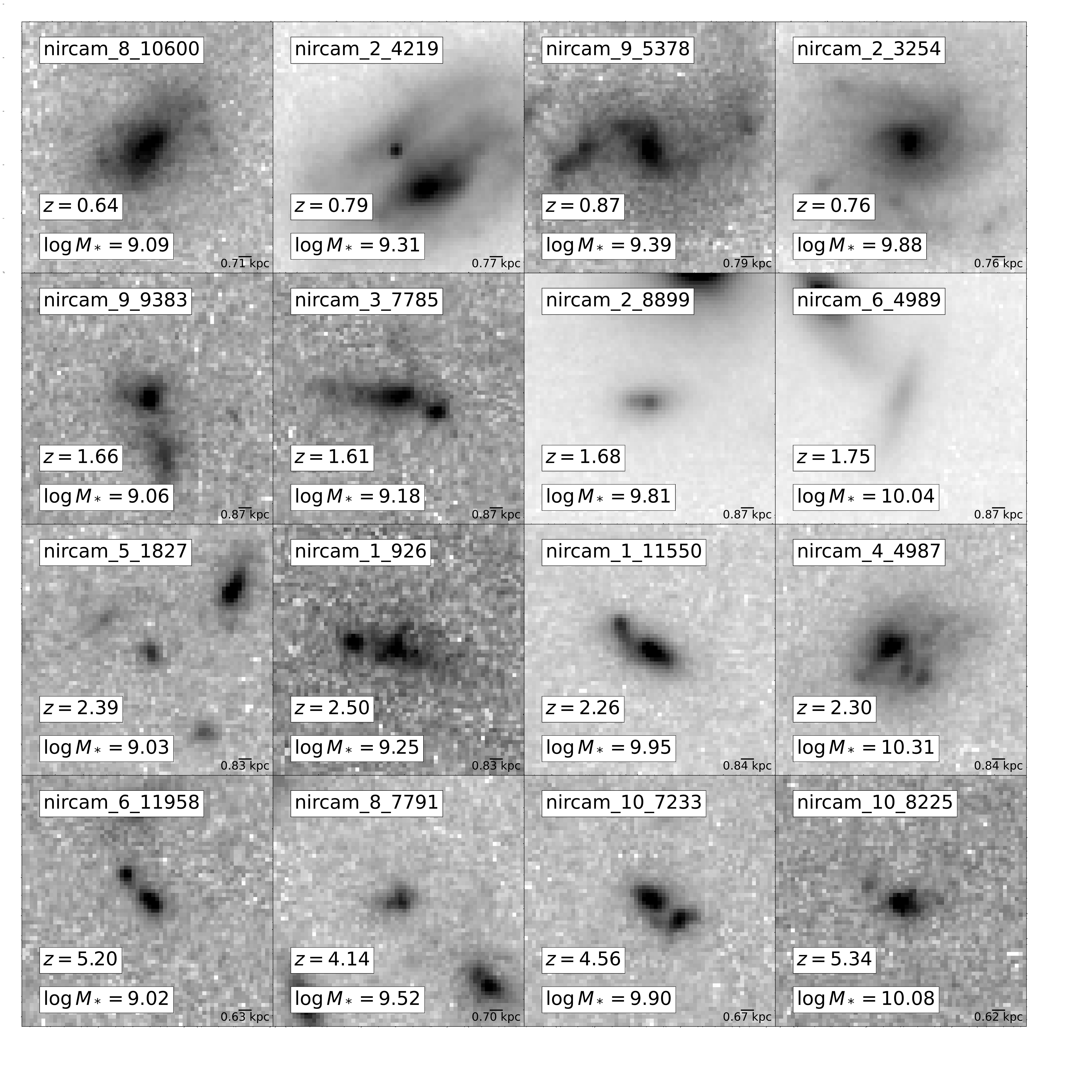}
    \end{minipage}%
    \begin{minipage}[t]{0.5\linewidth}
        \centering
        \captionsetup{font=footnotesize}
        \caption*{Bulge+Disk}
        \vspace{-10pt}
        \includegraphics[width=\linewidth]{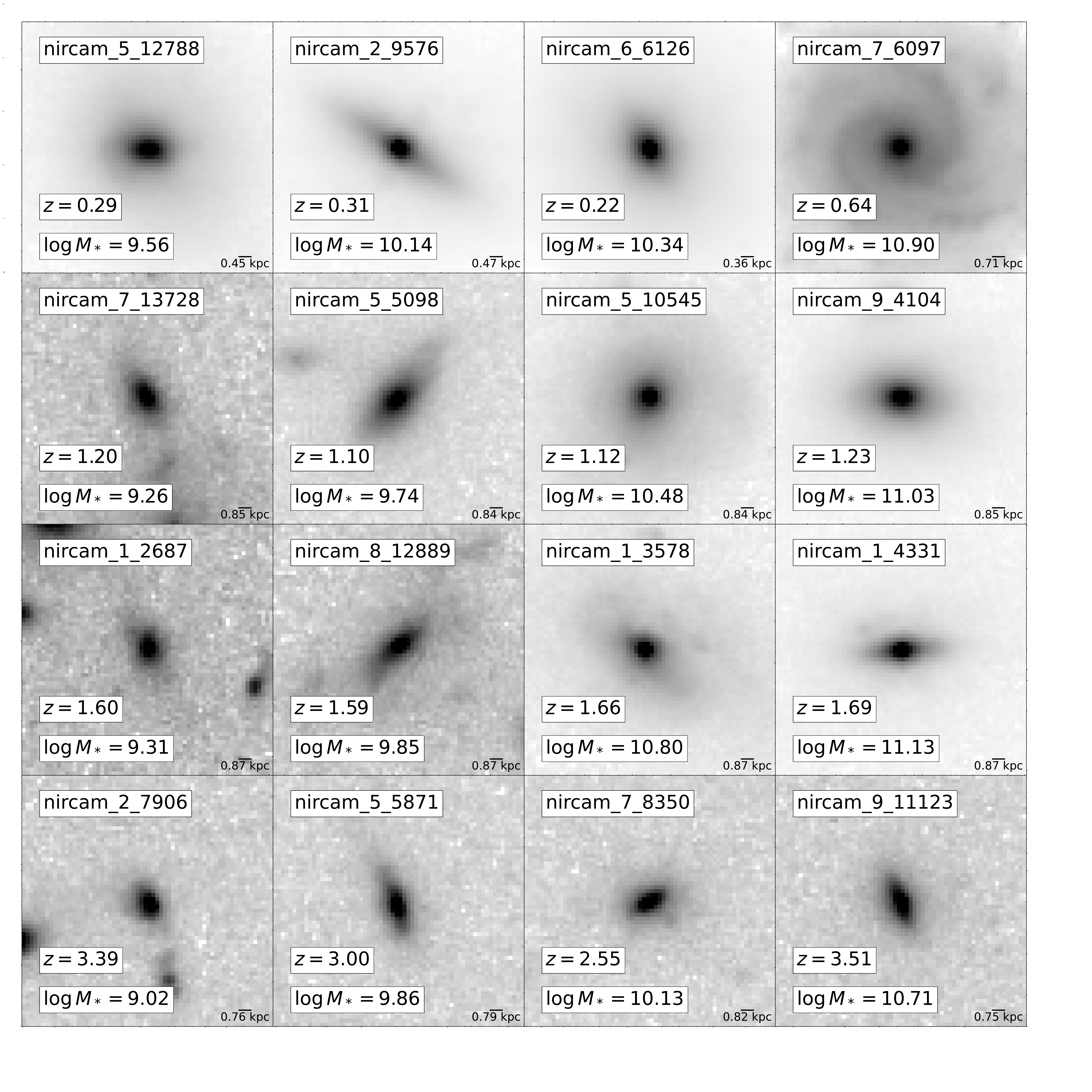}
    \end{minipage}
    
    \captionsetup{font=large}
    \caption{Example of random stamps of CEERS galaxies observed with the $F200W$ filter, classified in four main morphological classes. Each panel of 16 images illustrates a different class. Top left: spheroids, top right: disks, bottom left: irregulars, bottom right: composite bulge+disk galaxies. In each group of 16, galaxies are ordered by increasing photometric redshift (top to bottom) and stellar mass (left to right). The physical scale in kpc in shown for every galaxy. A square root scaling has been applied to enhance the outskirts.} 
    \label{fig:morph_visual_f200}
\end{figure*}

\begin{figure*}
    \begin{minipage}[t]{0.5\linewidth}
        \centering
        \captionsetup{font=footnotesize}
        \caption*{Spheroids}
        \vspace{-10pt}
        \includegraphics[width=\linewidth]{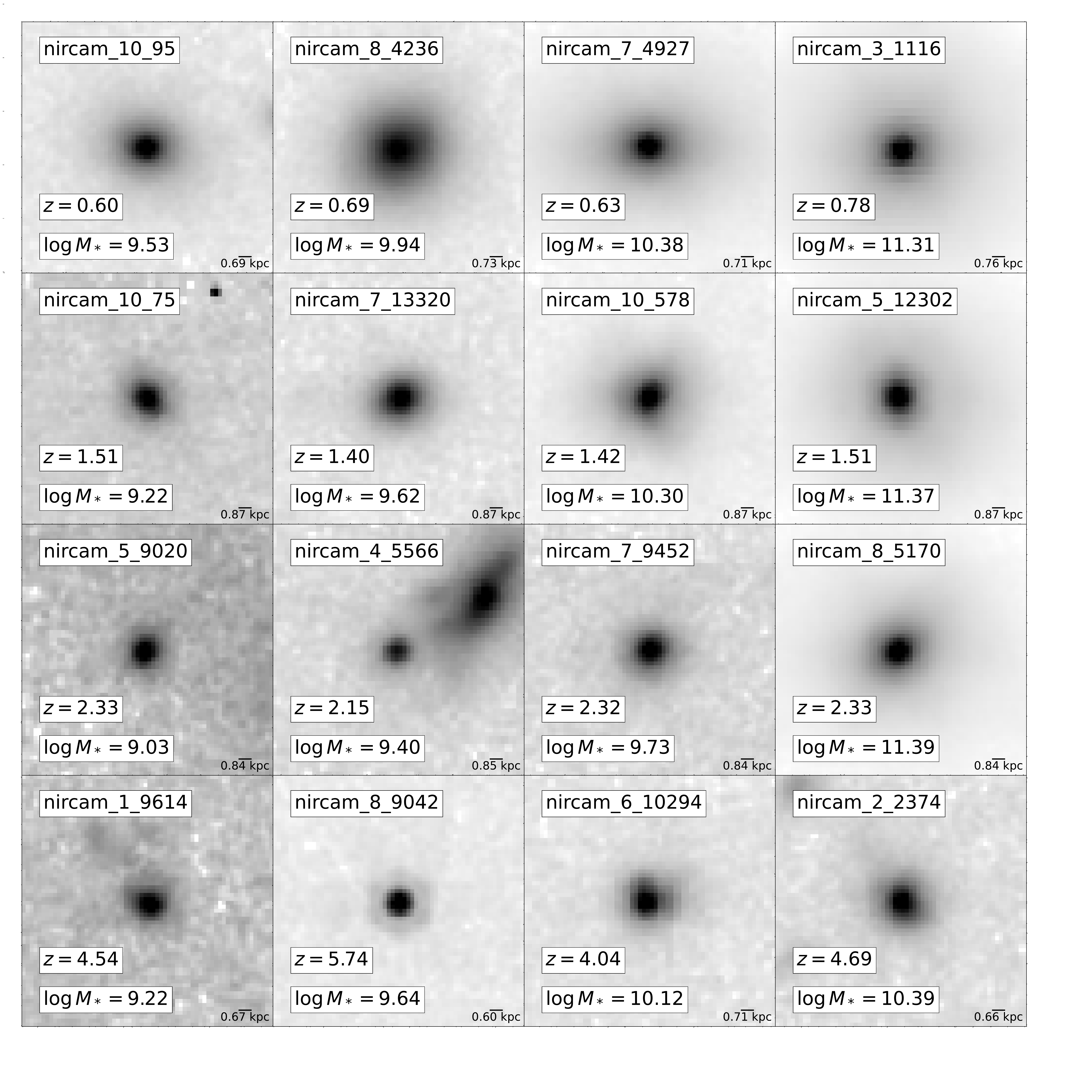}
    \end{minipage}%
    \begin{minipage}[t]{0.5\linewidth}
        \centering
        \captionsetup{font=footnotesize}
        \caption*{Disks}
        \vspace{-10pt}
        \includegraphics[width=\linewidth]{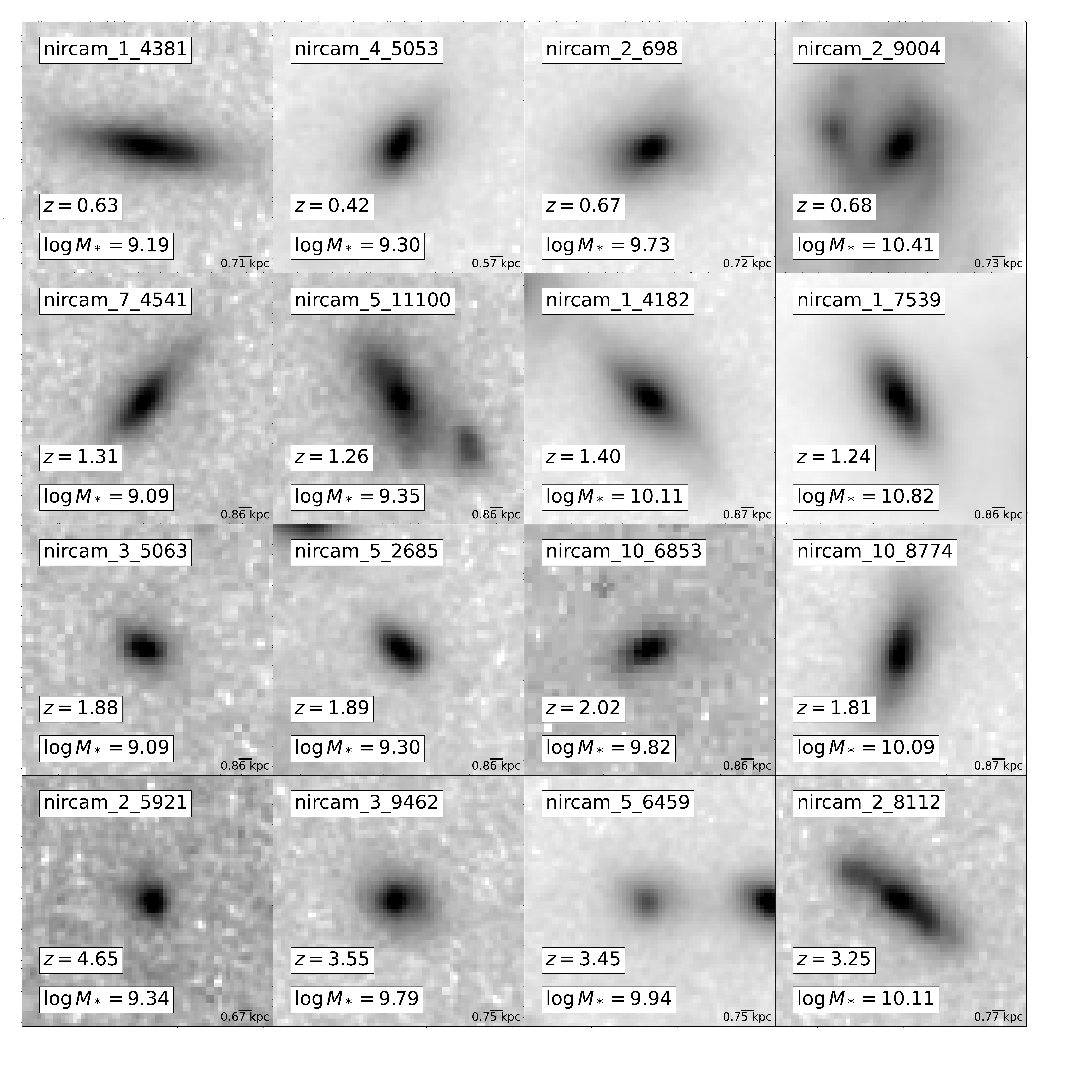}
    \end{minipage}
    
    \begin{minipage}[t]{0.5\linewidth}
        \centering
        \captionsetup{font=footnotesize}
        \caption*{Irregulars}
        \vspace{-10pt}
        \includegraphics[width=\linewidth]{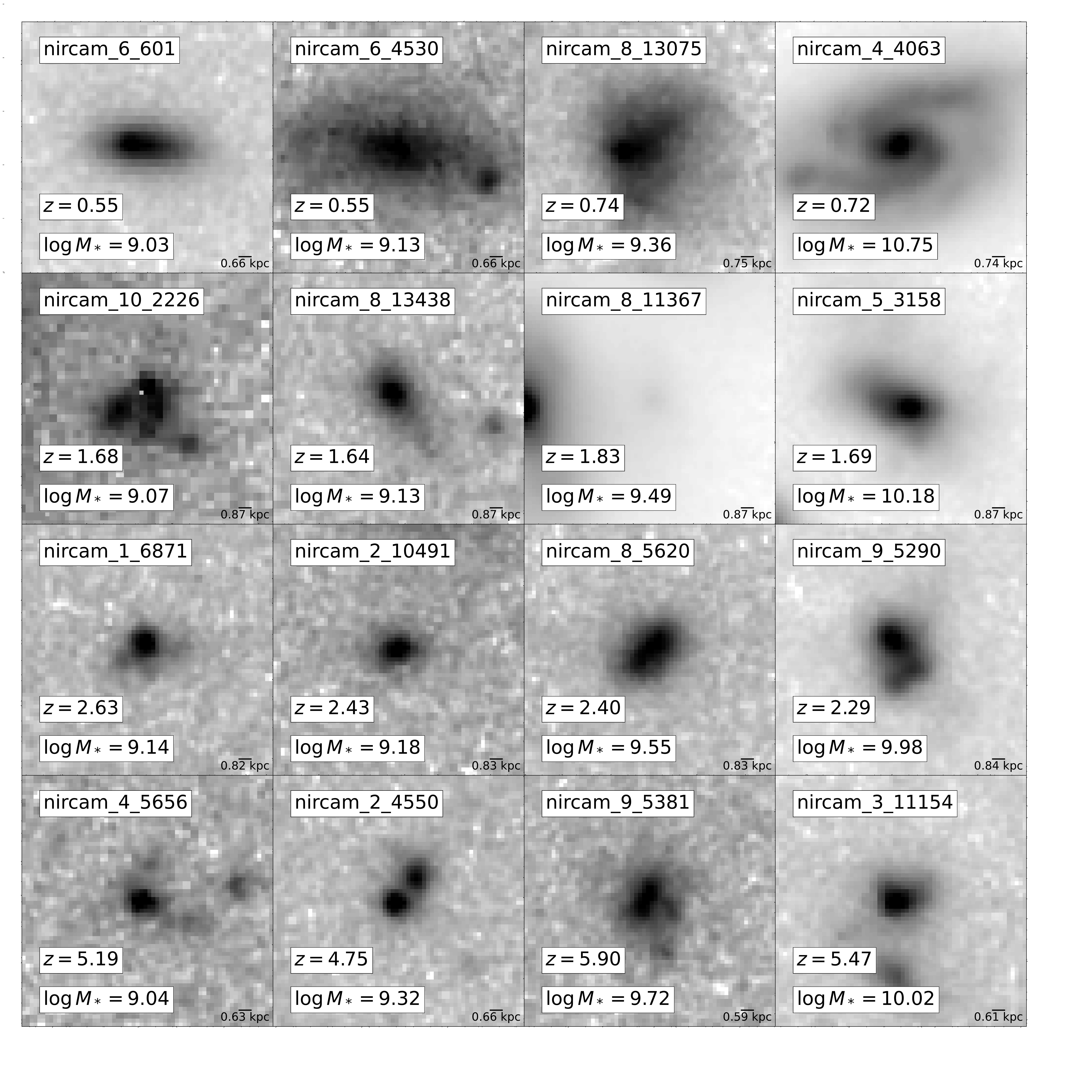}
    \end{minipage}%
    \begin{minipage}[t]{0.5\linewidth}
        \centering
        \captionsetup{font=footnotesize}
        \caption*{Bulge+Disk}
        \vspace{-10pt}
        \includegraphics[width=\linewidth]{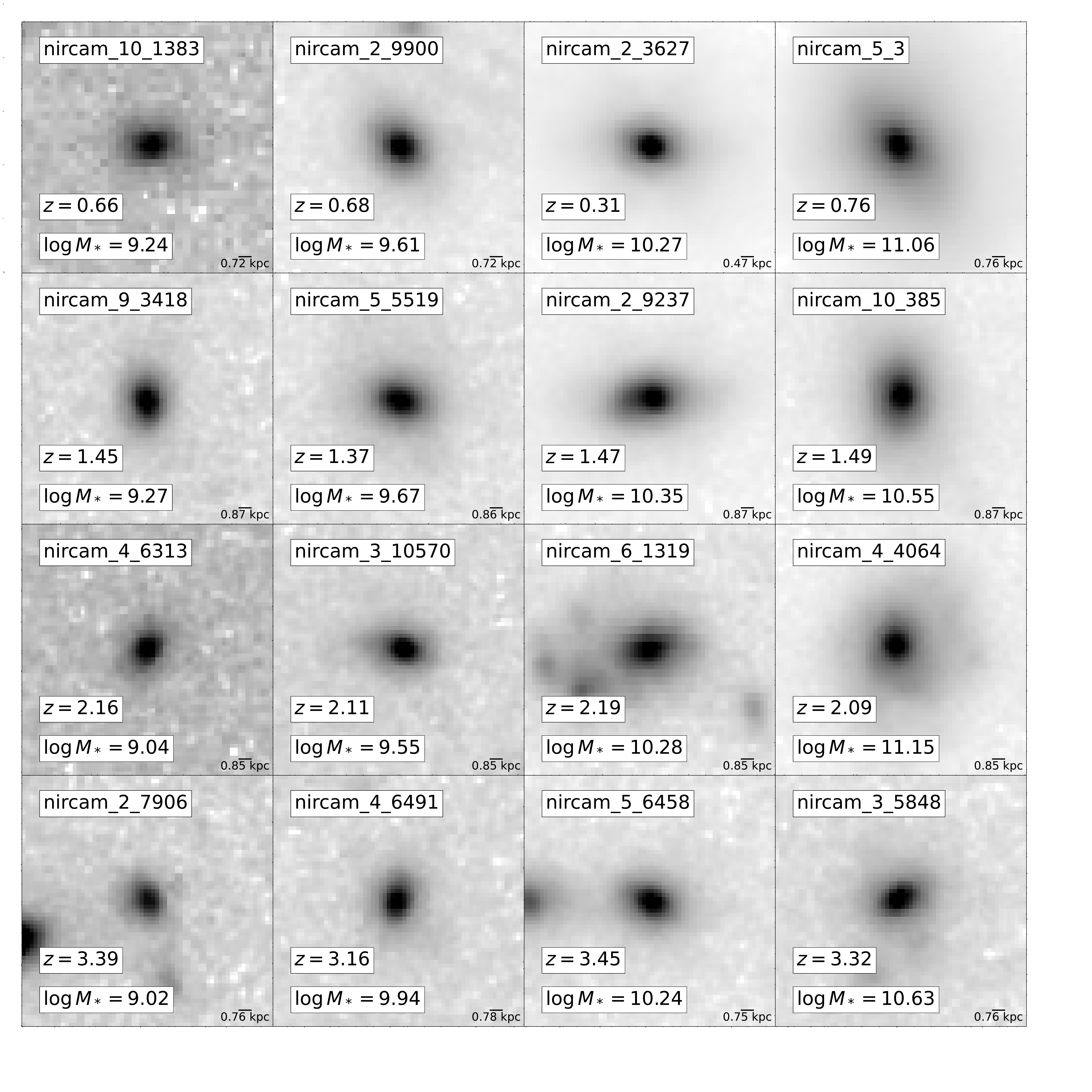}
    \end{minipage}
    
    \captionsetup{font=large}
    \caption{Example of random stamps of CEERS galaxies observed with the $F356W$ filter, classified in four main morphological classes. Each panel of 16 images illustrates a different class. Top left: spheroids, top right: disks, bottom left: irregulars, bottom right: composite bulge+disk galaxies. In each group of 16, galaxies are ordered by increasing photometric redshift (top to bottom) and stellar mass (left to right). The physical scale in kpc in shown for every galaxy. A square root scaling has been applied to enhance the outskirts.} 
    \label{fig:morph_visual_f356}
\end{figure*}

\begin{figure*}
    \begin{minipage}[t]{0.5\linewidth}
        \centering
        \captionsetup{font=footnotesize}
        \caption*{Spheroids}
        \vspace{-10pt}
        \includegraphics[width=\linewidth]{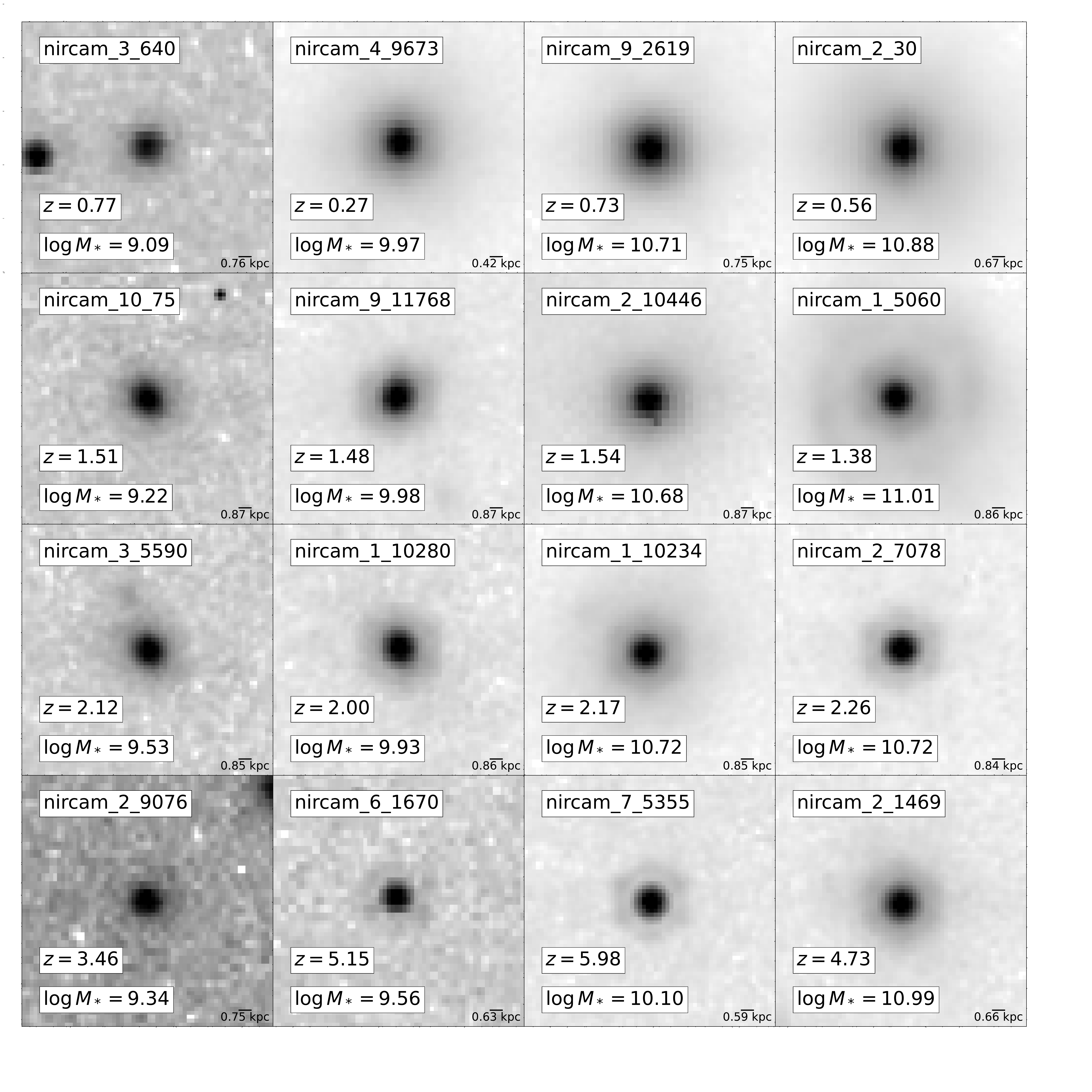}
    \end{minipage}%
    \begin{minipage}[t]{0.5\linewidth}
        \centering
        \captionsetup{font=footnotesize}
        \caption*{Disks}
        \vspace{-10pt}
        \includegraphics[width=\linewidth]{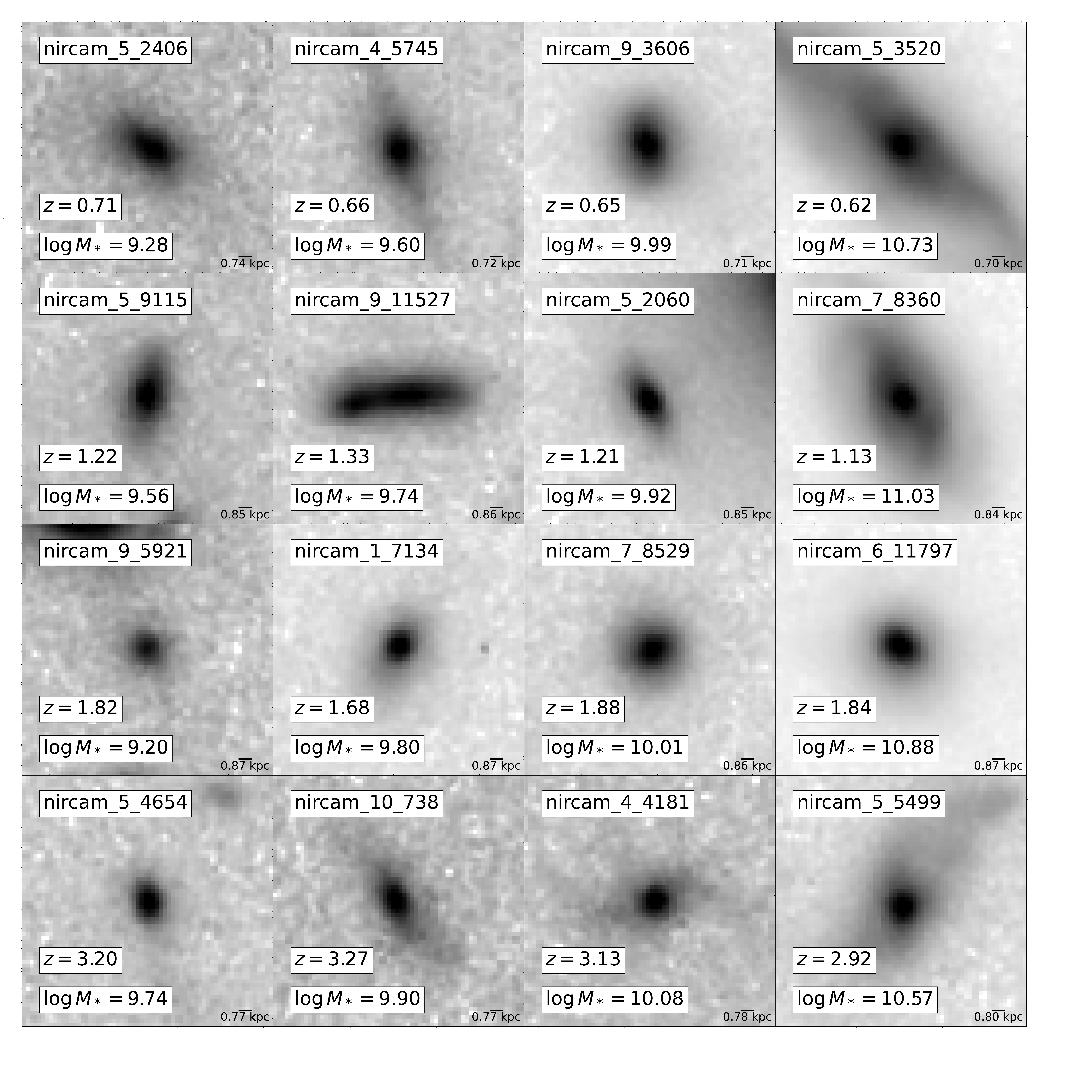}
    \end{minipage}
    
    \begin{minipage}[t]{0.5\linewidth}
        \centering
        \captionsetup{font=footnotesize}
        \caption*{Irregulars}
        \vspace{-10pt}
        \includegraphics[width=\linewidth]{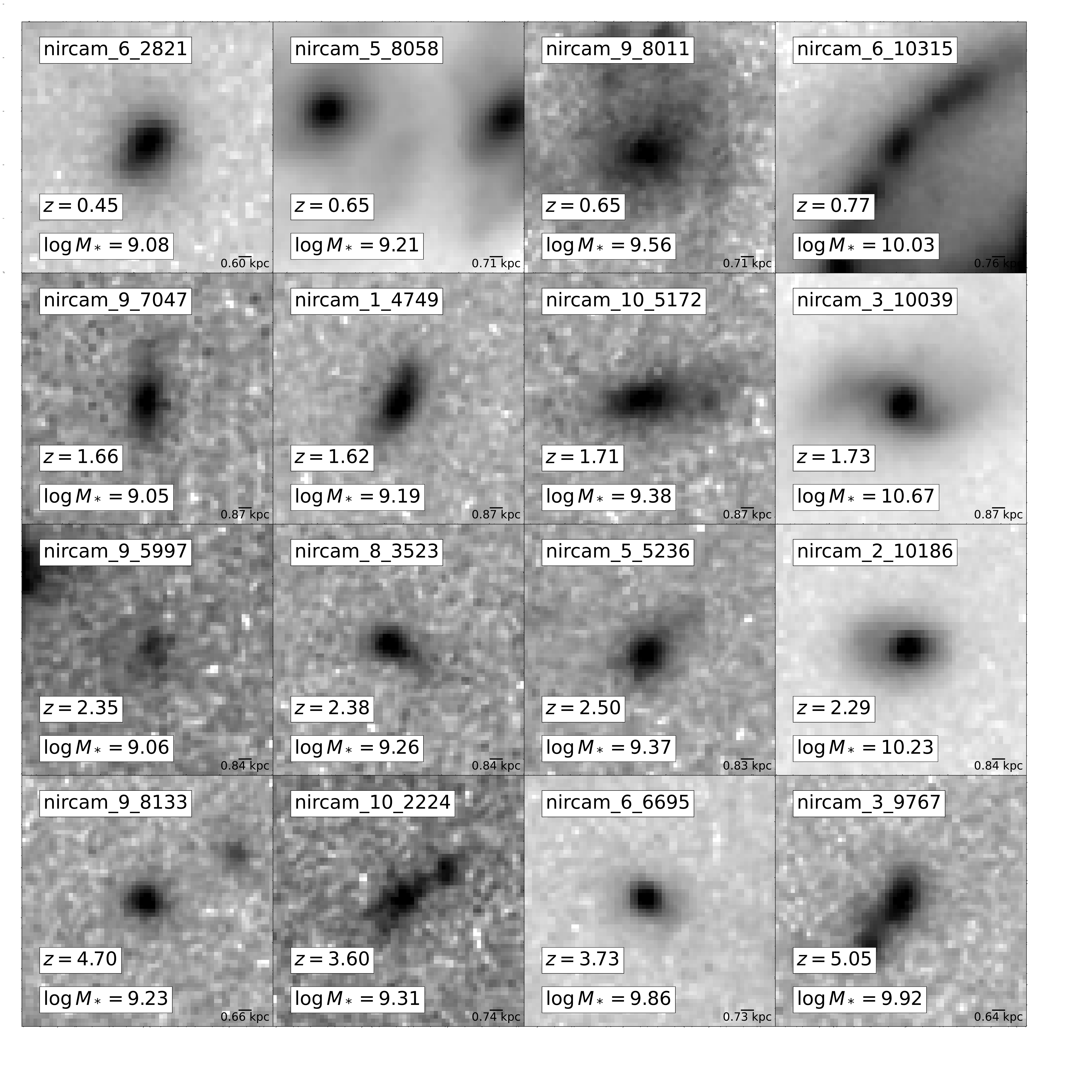}
    \end{minipage}%
    \begin{minipage}[t]{0.5\linewidth}
        \centering
        \captionsetup{font=footnotesize}
        \caption*{Bulge+Disk}
        \vspace{-10pt}
        \includegraphics[width=\linewidth]{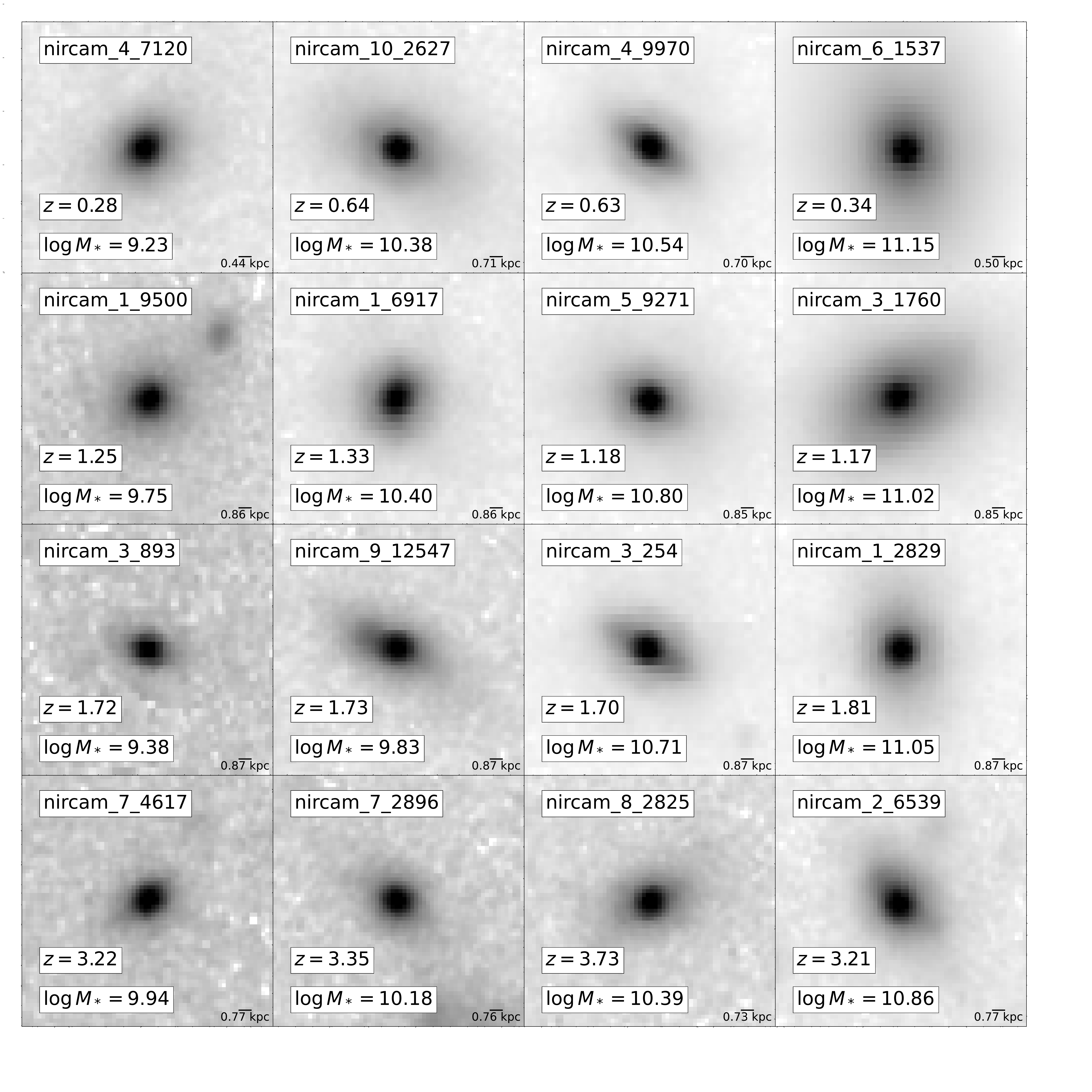}
    \end{minipage}
    
    \captionsetup{font=large}
    \caption{Example of random stamps of CEERS galaxies observed with the $F444W$ filter, classified in four main morphological classes. Each panel of 16 images illustrates a different class. Top left: spheroids, top right: disks, bottom left: irregulars, bottom right: composite bulge+disk galaxies. In each group of 16, galaxies are ordered by increasing photometric redshift (top to bottom) and stellar mass (left to right). The physical scale in kpc in shown for every galaxy. A square root scaling has been applied to enhance the outskirts.} 
    \label{fig:morph_visual_f444}
\end{figure*}

There are very few ground truth labels available for the CEERS images by construction. It is therefore not straightforward to assess the accuracy of the resulting morphological classification using the domain adaptation procedure described in the previous subsection. We followed two approaches to provide an overview of the quality of the resulting classification. First, we performed visual inspections of randomly selected galaxies belonging to the four classes.  Figures~\ref{fig:morph_visual_f200} to~\ref{fig:morph_visual_f444} show some random examples of galaxies ordered by stellar mass and redshift in the F200W, F356W, and F444W filters, classified in the four different classes\footnote{F200W, F356W, and F444W are the filters primarily used in the scientific analysis of this work and thus F150W images are not shown to keep a reasonably amount of figures. F150W images are used for comparison with HST as discussed in Section~\ref{sec:HST_JWST} and in the Appendix~\ref{sec:hst_jwst}.}. The stamps clearly show distinct morphological features, indicating that the network produces meaningful classifications on JWST images, even without provided labels. The difference between the two main classes - spheroids and disks - shown in the top row of figures~\ref{fig:morph_visual_f200} to~\ref{fig:morph_visual_f444} is obvious and shows, not surprisingly, that the network is clearly able to distinguish between compact and extended sources. The boundary between disks and irregulars is slightly more diffuse - but this is known to be a difficult task. The figure seems to suggest that irregular galaxies might be biased towards images that contain multiple objects in the field of view. Some look indeed as perturbed light profiles but others might simply be foreground/background contamination. It is also worth noticing the different spatial resolution of the different filters. Some compact galaxies appear to be unresolved in the redder wavebands.

\subsection{Comparison with visual classifications}

\begin{figure*}
    {\includegraphics[width=0.5\linewidth]{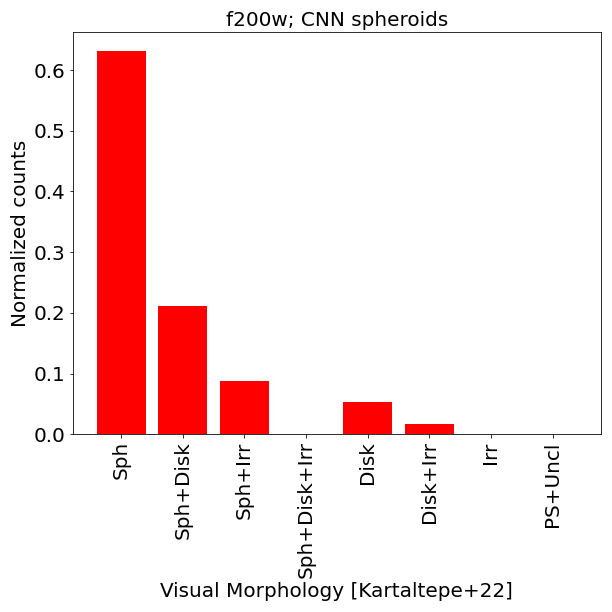}}
    {\includegraphics[width=0.5\linewidth]{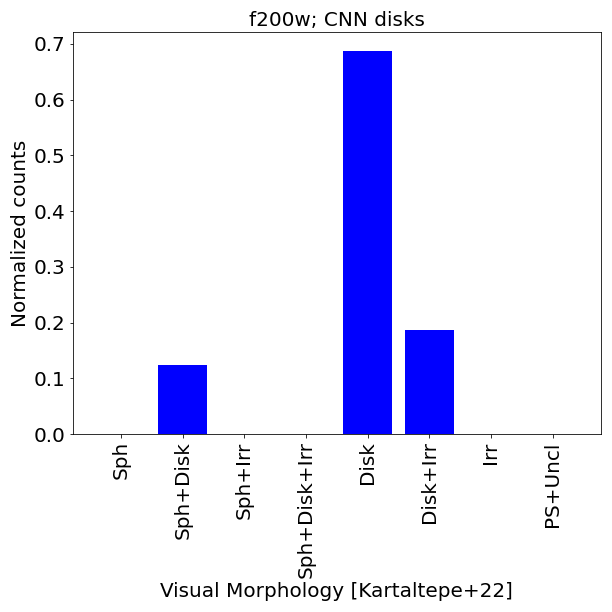}}
   {\includegraphics[width=0.5\linewidth]{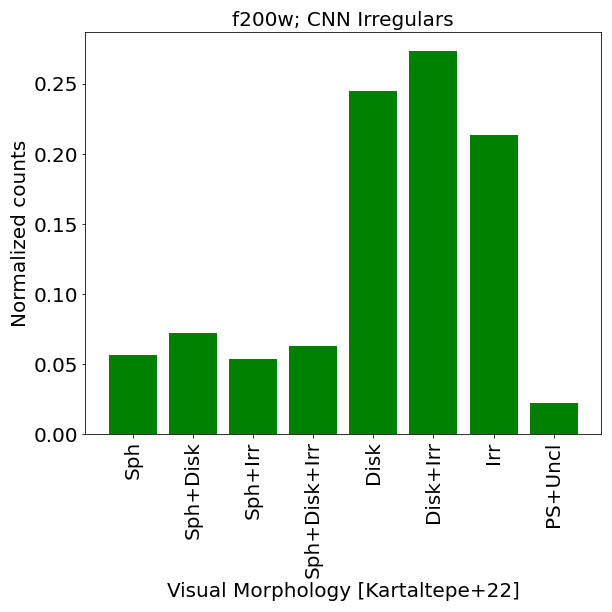}}
    {\includegraphics[width=0.5\linewidth]{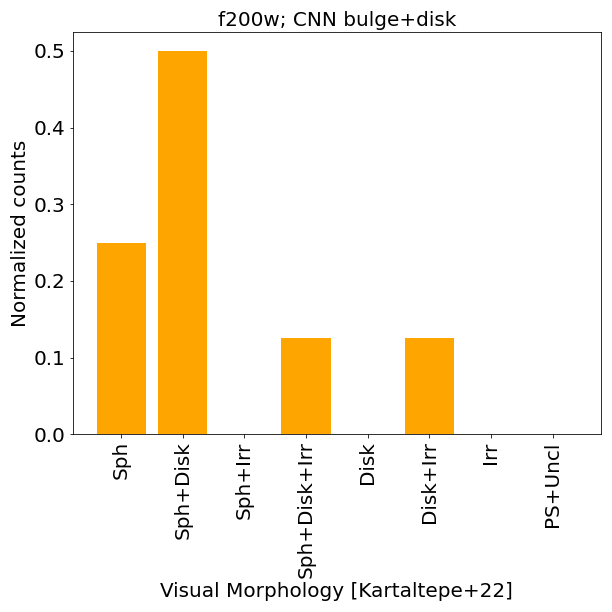}}
    \caption{Comparison between the deep learning based classification in the $F200W$ band presented in this work and the visual classifications of~\cite{2022arXiv221014713K} for galaxies with $z>3$. The different panels show the distribution of visual classes for each of the four classes defined in this work; top left: spheroids, top right: disks, bottom left: irregulars, bottom right: disks+spheroids. } 
    \label{fig:morph_jeyhan}
\end{figure*}

In a recent study, \cite{2022arXiv221014713K} visually classified a subset of galaxies with $z > 3$ using a similar scheme to the one used for CANDELS. As an independent cross-check of the domain-adapted morphologies, we compare our deep learning-based classification to the visual one in~\autoref{fig:morph_jeyhan}. For this comparison, we use the neural network-based classification in the $F200W$ filter, since it is the primary band also used in~\cite{2022arXiv221014713K}, although classifiers could also inspect other bands. Although this comparison is performed on a reduced dataset of $\sim 800$ galaxies, it is crucial since it compares two independent classifications made on the same images. It is also the set of galaxies for which more discrepancies might be expected, as they represent the faint end of the distribution and cover a redshift range not well probed by HST.

Since the classification of~\cite{2022arXiv221014713K} is more detailed than ours, figure~\ref{fig:morph_jeyhan} shows, for each one of the four deep learning classes, the distribution of visual labels as defined in~\cite{2022arXiv221014713K}. Overall, we observe good agreement between the two classifications. The distributions for the four primary classes are clearly different. More than 90\% of galaxies classified as spheroids by the neural network are also flagged as having a spheroid component visually. All galaxies automatically classified as disks have visually identified disks. Even the class of composite bulge + disk systems shows good agreement with visual classifications, with a clear peak at the Sph+Disk systems. As hinted by the visual inspection of the previous section, the largest discrepancies are between the irregular and disk classes. About 25\% of galaxies classified as disks by the neural networks are flagged as irregulars by human classifiers, and the same happens in the other direction. Given the difficulty - and somewhat subjective nature - of identifying irregular galaxies and the fact that the two classifications compared here are completely independent, we consider this contamination acceptable.

\section{From HST to JWST: the impact of depth and spatial resolution}

\label{sec:HST_JWST}

JWST images provide deeper and higher spatial resolution images compared to the HST, which until recently was the primary telescope used to quantify galaxy morphologies. The quality of morphological classification is known to be significantly affected by factors such as resolution and signal-to-noise ratio. Hence, it is interesting to compare the classifications of the same galaxies observed with HST and JWST. This is the main focus of this subsection, and to minimize the wavelength effect, we compare the $F160W$ and $F150W$ images.

We begin by exploring in Figures~\ref{fig:conf_mat_CEERS_CANDELS} and~\ref{fig:hex_CEERS_CANDELS} how basic morphological classifications on the same objects change between JWST and HST. For simplicity and to better understand the differences between the two classifications, we consider two cases: early vs. late type and disturbed vs. undisturbed.

In the early-type class, we include both spheroids and bulge+disk galaxies as defined in section~\ref{sec:data}.~\cite{2015ApJS..221....8H} showed that the bulge+disk class is mainly composed of bulge-dominated systems. The late-type class contains both disks and irregulars.

The disturbed class contains irregulars, while the undisturbed class gathers all the remaining three classes (see section~\ref{sec:data}).

Since the primary differences between the two telescopes are sensitivity and spatial resolution, in addition to a global comparison, we quantify the differences as a function of apparent $F150W$ magnitude as a proxy for signal-to-noise ratio and apparent half-light size as a proxy for resolution.

\subsection{Early/late type galaxies}

\begin{figure}
    
{\includegraphics[width=\linewidth]{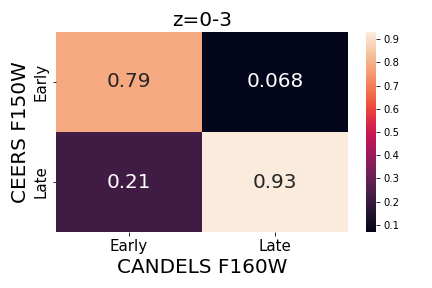}}
{\includegraphics[width=\linewidth]{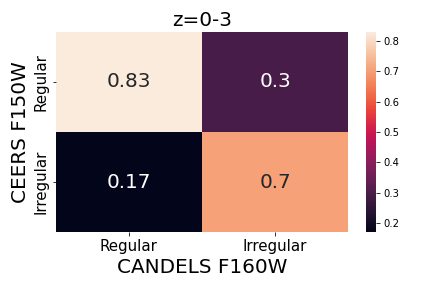}}

    \caption{Confusion matrices showing the overall agreement between early/late (top) and regular/irregular (bottom) classifications with HST-WFC3 F160W imaging and JWST-NIRCam F150W imaging for galaxies in the redshift range $z=0-3$. } 
    \label{fig:conf_mat_CEERS_CANDELS}
\end{figure}

The top panel of Figure~\ref{fig:conf_mat_CEERS_CANDELS} shows that, overall, there is a good agreement between HST and JWST based early/late classifications. We find that $\sim90\%$ of the galaxies between $z=0-3$ have the same classifications with both instruments. The confusion matrix also shows that the agreement is not completely symmetrical, i.e. $\sim20\%$ of galaxies classified as early-type with HST move to the late-type class with JWST, in agreement with previous results suggesting that disks are more abundant than expected~\citep{2022ApJ...938L...2F}. In Figure~\ref{fig:hex_CEERS_CANDELS} we explore into more details how the differences depend on apparent magnitude and apparent size. The discrepancy in the relative fraction of early type galaxies between JWST and HST ($\Delta_{early}= \frac{N_{early}^{CEERS}-N_{early}^{CANDELS}}{N_{total}}$) is consistently below $\sim10\%$ across most of the parameter space. Only for very small galaxies ($R_e<0.1^{"}$) does the relative fraction of early-type galaxies measured by JWST drop by about $\sim30\%$ compared to HST. This increase in discrepancy for small and faint galaxies suggests that it is related to spatial resolution and possibly signal-to-noise. We show in the Appendix~\ref{sec:hst_jwst} some example stamps of galaxies with different morphological classifications in CANDELS and CEERS.  Interestingly, because of the distribution of galaxies in the parameter space (as shown in the left panel of figure~\ref{fig:hex_CEERS_CANDELS}), these discrepancies only affect a small fraction of objects. Thus, the measured fractions of early and late-type objects as a function of redshift and stellar mass (shown in the top row of figure~\ref{fig:frac_CEERS_CANDELS}) are very similar and fully compatible within the $1\sigma$ uncertainties.

\subsection{Peculiar galaxies}

\begin{figure*}
    
{\includegraphics[width=0.33\linewidth]{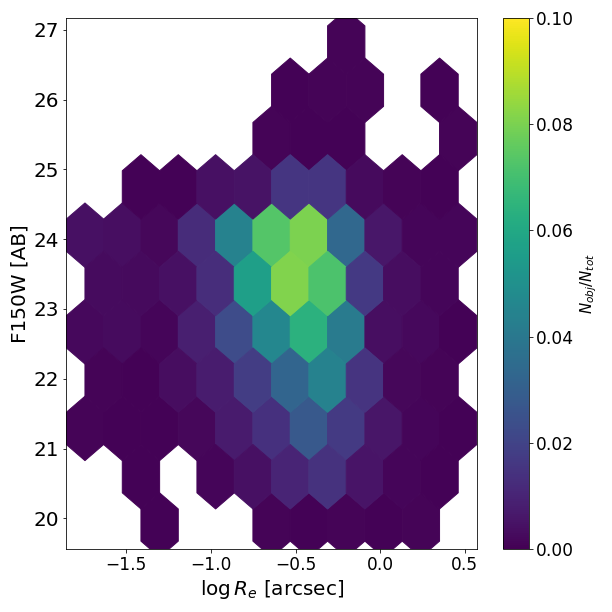}}
{\includegraphics[width=0.33\linewidth]{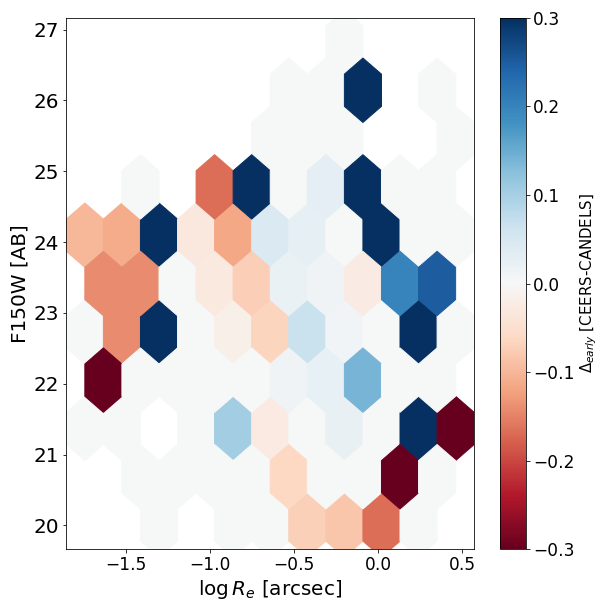}} {\includegraphics[width=0.33\linewidth]{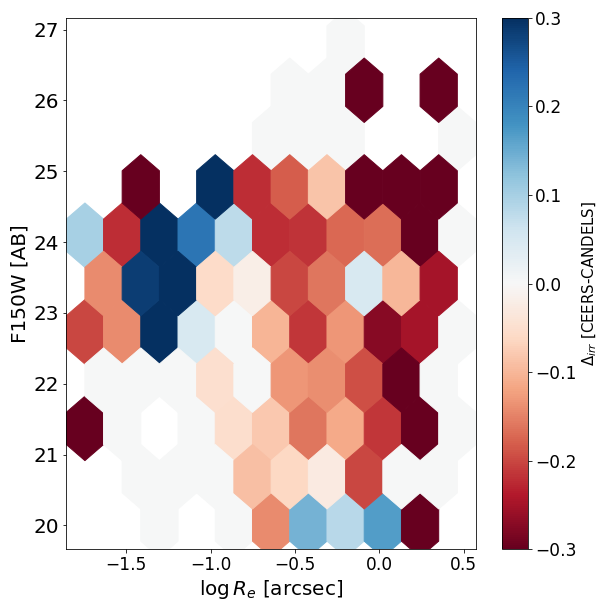}}

    \caption{Differences between CANDELS and CEERS measured morphologies as a function of apparent $F150W$ magnitude and angular half-light radius. Left panel: Number density of objects. Middle panel: Difference between the fraction of early-type galaxies in CEERS and CANDELS: $\Delta_{early}= \frac{N_{early}^{CEERS}-N_{early}^{CANDELS}}{N_{total}}$. Right panel: Difference between the fraction of disturbed galaxies in CEERS and CANDELS: $\Delta_{irr}= \frac{N_{irr}^{CEERS}-N_{irr}^{CANDELS}}{N_{total}}$. } 
    \label{fig:hex_CEERS_CANDELS}
\end{figure*}

\begin{figure*}
    
{\includegraphics[width=\linewidth]{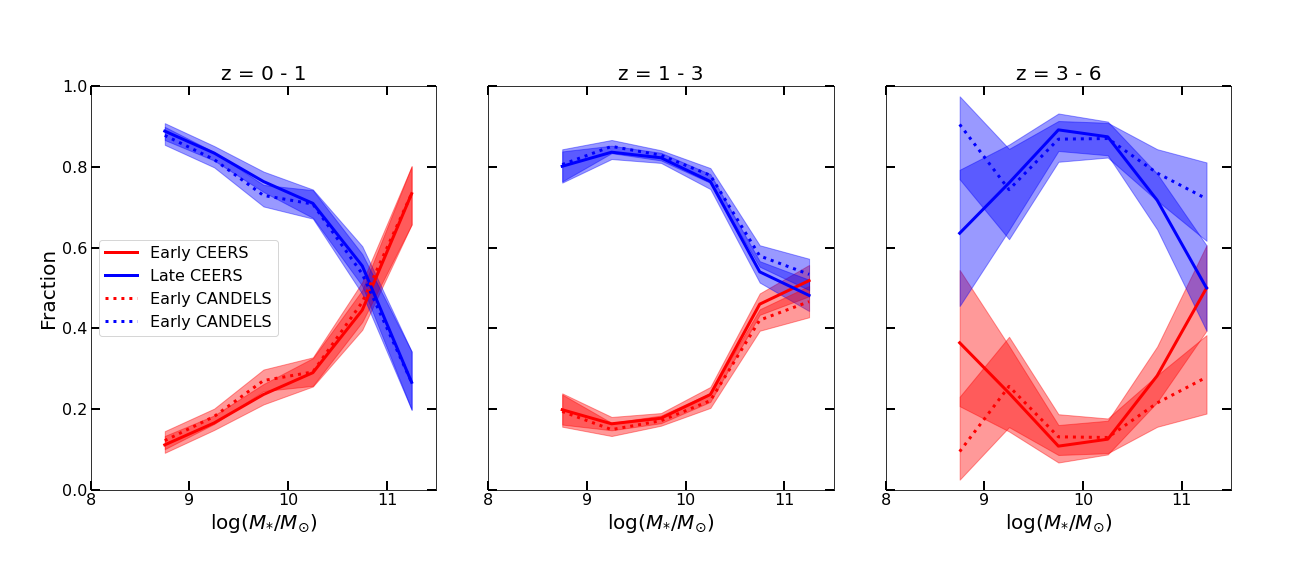}}\vspace{-1cm}

{\includegraphics[width=\linewidth]{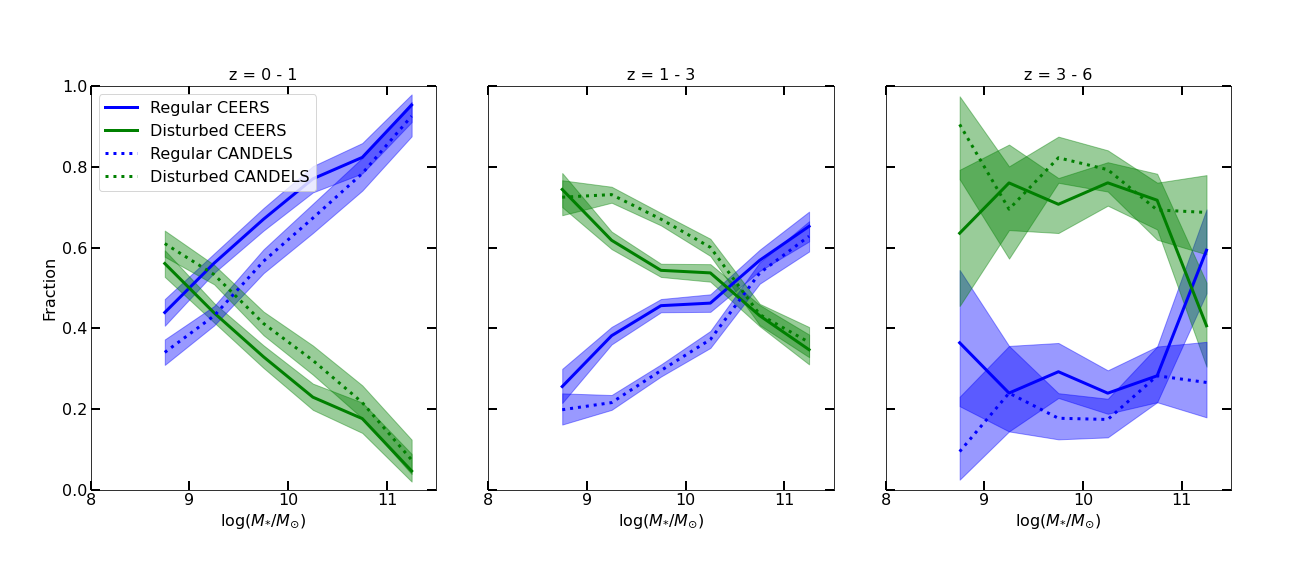}}
    
    \caption{Comparison between the fractions of early and late-type galaxies (top row) and disturbed and undisturbed galaxies (bottom row) as a function of stellar mass measured in CANDELS (dashed lines)  and CEERS (sold lines) for exactly the same galaxies. For CANDELS, the morphologies are inferred in the $F160W$ filter while for CEERS we use $F150W$ in all panels to reduce wavelength induced differences in the morphologies. Each panel indicates a redshift bin as labeled. The shaded regions show Poisson uncertainties.} 
    \label{fig:frac_CEERS_CANDELS}
\end{figure*}

The differences between the fractions of disturbed and undisturbed galaxies are more significant. The top panel of Figure~\ref{fig:conf_mat_CEERS_CANDELS} shows that the overall agreement is of $\sim76\%$. This is mostly driven by the fact that $\sim30\%$ of galaxies classified as irregular with HST are not found disturbed with JWST. The rightmost panel of figure~\ref{fig:hex_CEERS_CANDELS} confirms a clear trend of classifying galaxies as less disturbed in JWST than in HST images. The differences in the relative fractions of disturbed systems ($\Delta_{irr}= \frac{N_{irr}^{CEERS}-N_{irr}^{CANDELS}}{N_{total}}$) range on average between $\sim20$ and $\sim40\%$ and tend to be more pronounced for large objects. This trend is likely due to the fact that the disturbed/irregular class is a poorly defined category of objects that includes all galaxies with some irregularity in their surface brightness profile. Thus, the irregular appearance is highly dependent on the signal-to-noise ratio since fluctuations in the surface brightness caused by noise can be easily interpreted as an irregular light distribution. Moreover, deeper observations will tend to better detect a potential diffuse component around galaxies which can be interpreted as a disk and move the classification from irregular to disk. Given that the CEERS survey's imaging is significantly deeper than the CANDELS data, it is expected that galaxies will appear less disturbed. This trend is confirmed by the rightmost panel of figure~\ref{fig:hex_CEERS_CANDELS}. Extended galaxies have lower SNR per pixel at a fixed magnitude, which may explain why the differences are larger for larger objects. One could argue as well that deeper observation can allow one to better detect low surface brightness features around galaxies which might make them look more irregular and therefore have the opposite effect. This is not what Figure~\ref{fig:hex_CEERS_CANDELS} suggests, and the reason might be that the JWST observations, although deep, are not deep enough to detect these low surface brightness features which are typically seen at depths of $\sim30$ $mag.arcsec^{-2}$ or more (e.g.,~\citealp{2016ApJ...823..123T}). Some examples of images with different classifications are shown in the Appendix~\ref{sec:hst_jwst}. Despite these differences, the impact on the measured fraction of disturbed galaxies as a function of stellar mass and redshift reported in figure~\ref{fig:frac_CEERS_CANDELS} is not significant. The fraction of irregulars tends to be less than $\sim10\%$ smaller in CEERS than in CANDELS. This can be explained by the fact that the majority of objects lie in a region of the parameter space where the differences between HST and JWST are less pronounced. We also notice that the leftmost panel of Figure~\ref{fig:frac_CEERS_CANDELS} shows that irregular galaxies completely dominate the galaxy distribution at $z>3$. This is most likely a consequence of using the $F200W$ filter which probes the UV rest-frame as discussed in the following. It might be also affected by selection biases since we are only representing objects in common in CANDELS and CEERS and that's also why it differs from the results of~\cite{2022arXiv221014713K}.



\section{Galaxy morphology since $z=6$ in the rest-frame NIR}
\label{sec:morph_evol}
A unique feature of JWST is that it probes longer wavelengths than HST, offering a view of galaxy structure in the mid-infrared. This enables to probe galaxy structure in the optical rest-frame up to higher redshifts ($z\sim 6$) but also in the NIR rest-frame with a proxy closer to stellar mass and less biased by stellar age and/or dust light-weighting effects.  

\subsection{Morphology vs. Wavelength}

\begin{figure*}
    
{\includegraphics[width=0.5\linewidth]{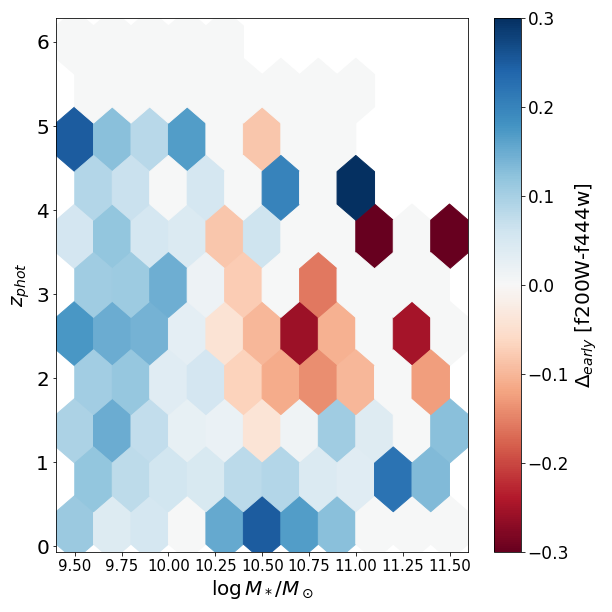}}
{\includegraphics[width=0.5\linewidth]{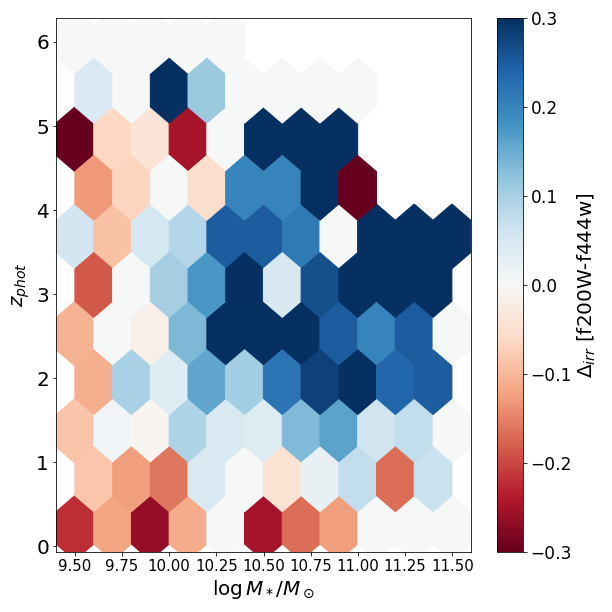}}

    \caption{Differences between morphologies measured in F200W and F444W in CEERS. The left panel shows the difference in the fractions of early-type galaxies ($\Delta_{early}= \frac{N_{early}^{F200w}-N_{early}^{F444w}}{N_{total}}$) and the right panel differences in the fractions of disturbed galaxies ($\Delta_{irr}= \frac{N_{irr}^{F200w}-N_{irr}^{F444w}}{N_{total}}$). } 
    \label{fig:hexbin_f200_f444}
\end{figure*}

In a similar fashion to the analysis of the previous section, we investigate how galaxy morphology changes with wavelength. Figure~\ref{fig:hexbin_f200_f444} compares the fractions of early-type and peculiar galaxies between the $F200W$ and $F444W$ filters as a function of stellar mass and redshift.

\begin{figure*}
    
{\includegraphics[width=\linewidth]{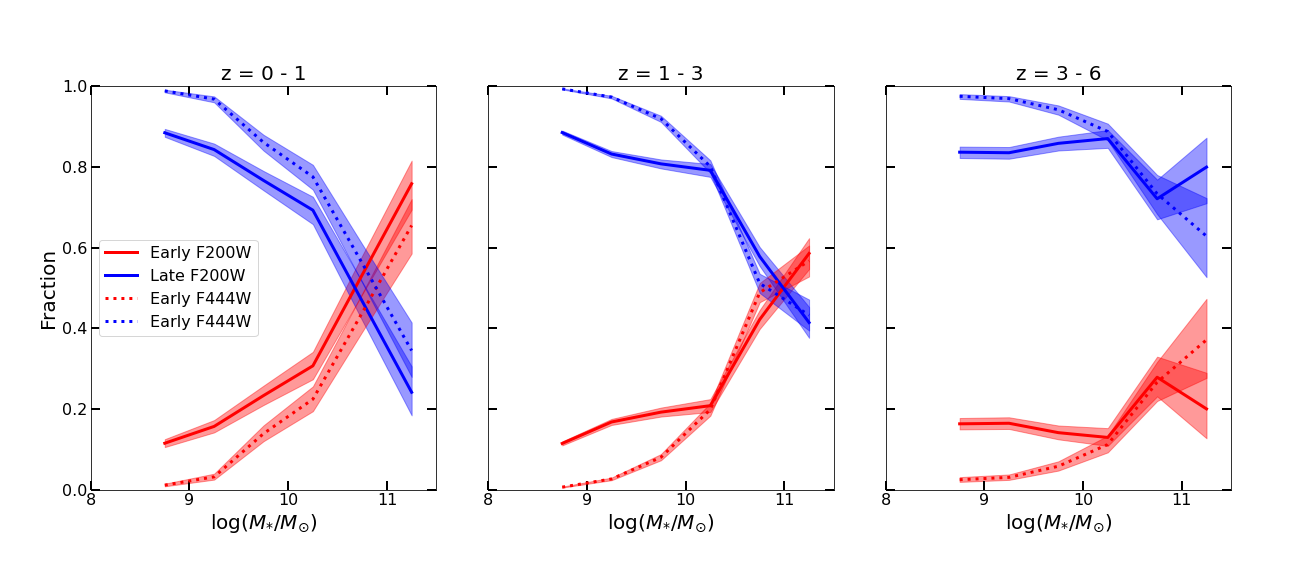}}\vspace{-1cm}

{\includegraphics[width=\linewidth]{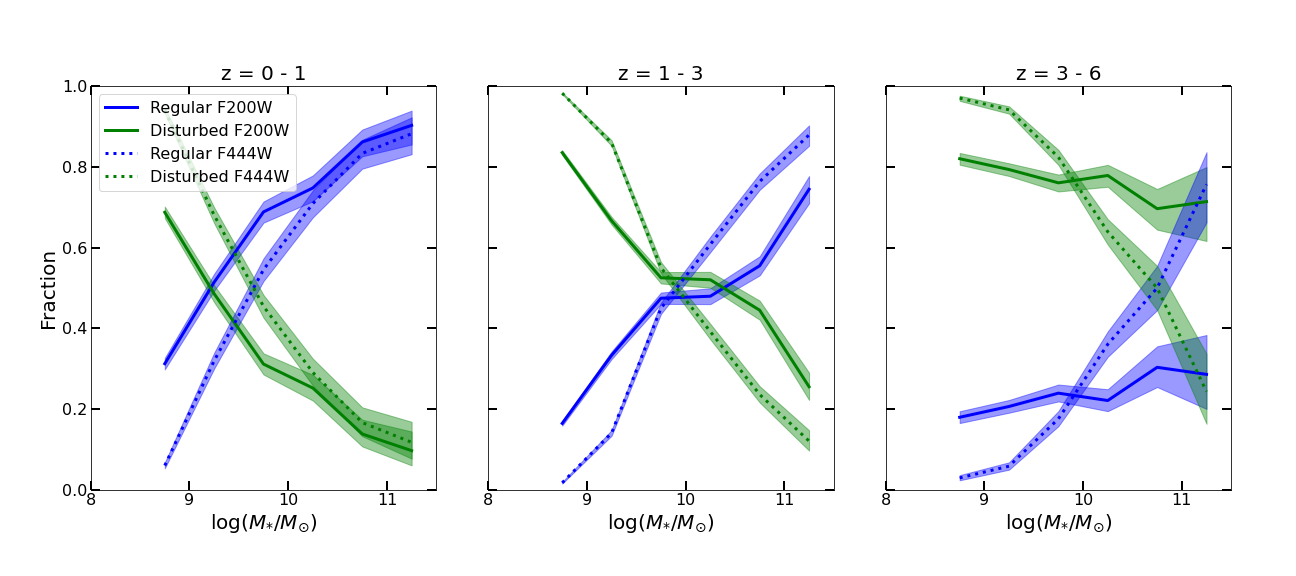}}

    \caption{Comparison of the morphological fractions as function of redshift and stellar mass measured in the $F200W$ and $F444W$ filters. The top row show the fractions of early (red lines) an late (blue) type galaxies in three redshift bins as labeled in the $F200W$ (solid lines) and $F444W$ (dashed lines) bands. The bottom row, indicates the fraction of regular (blue) and peculiar (green) galaxies.  } 
    \label{fig:morph_fractions_444_200}
\end{figure*}

Concerning the abundances of early-type galaxies, the figure shows that the differences between the fractions measured in the two filters become significant for galaxies above $z\sim3$ and stellar masses larger than $\log M_*\sim10.5$. In particular, we find that the fraction of early-type galaxies is $30\%$ larger in the longer wavelengths. This result is expected since, at $z>3$, the $F200W$ filter probes the near-UV and is therefore more sensitive to the emission of young stars, which can make the bulge difficult to detect (e.g.,~\citealp{2023arXiv230312845P}). The $F444W$ filter, on the other hand, probes a rest-frame wavelength between $\sim1$ and $\sim0.8$ $\mu m$ above $z\sim3$, which is dominated by old stars typically located in the central parts of galaxies (i.e., bulge). For stellar masses lower than $10^{10}$ solar masses, the differences between the two filters are slightly less pronounced, and the trend is inverted. Specifically, the fraction of early-type galaxies is larger in the bluer filters. This counterintuitive trend may reflect the presence of compact star-forming regions interpreted as bulge components in the $F200W$ filter. It may also be a resolution effect which prevents the disk component to be detected in the longer wavelengths. We discuss this further in section~\ref{sec:morph_evol}. The top row of Figure~\ref{fig:morph_fractions_444_200} explores how these reported differences between the two filters translate into the evolution of early and late-type fractions as a function of stellar mass and redshift. We observe the expected behavior from the trends described in Figure~\ref{fig:hexbin_f200_f444}. The fraction of early-type galaxies in the highest redshift bin slightly increases at the high-mass end in the $F444W$ filter, while the opposite is measured at the low-mass end.

The differences between the fractions of disturbed and undisturbed galaxies are more pronounced. At $z>3$ the fraction of irregular galaxies in the $F444W$ filter can be up to $\sim50\%$ smaller than in the $F200W$. This might be a signature that the distribution of mass is less irregular than the one of light, as pointed out by previous works who attempted to estimate stellar mass maps from HST imaging (e.g.,\citealp{2012ApJ...753..114W}). It is thus interesting to see that this trend is confirmed when probing galaxy morphology at longer wavelengths. It is worth noticing however that a similar trend could be driven by a difference in spatial resolution. The long wavelength imaging has a factor of $\sim2$ lower resolution than the short wavelength (see figures~\ref{fig:morph_visual_f200} and~\ref{fig:morph_visual_f444}) so it is less sensitive to substructure, which could make galaxies appear more regular in their light distribution. However, in Section~\ref{sec:HST_JWST} we report that the fraction of peculiar galaxies decreases in JWST imaging as compared to HST, even with increased spatial resolution. The trend measured with wavelength hence suggests that it is not purely a resolution effect. The bottom panel of Figure~\ref{fig:morph_fractions_444_200} measures how these differences between the filters is translated into the fraction of peculiar galaxies as a function of stellar mass and redshift. The differences are particularly dramatic at $z>3$. The fraction of peculiar galaxies completely dominates in the $F200W$ band at all stellar masses. In the $F444W$ filter though, they only dominate at the low mass end.

\subsubsection{Morphological evolution}

\begin{figure*}
    
\centering

\subfigure{\includegraphics[width=\linewidth]{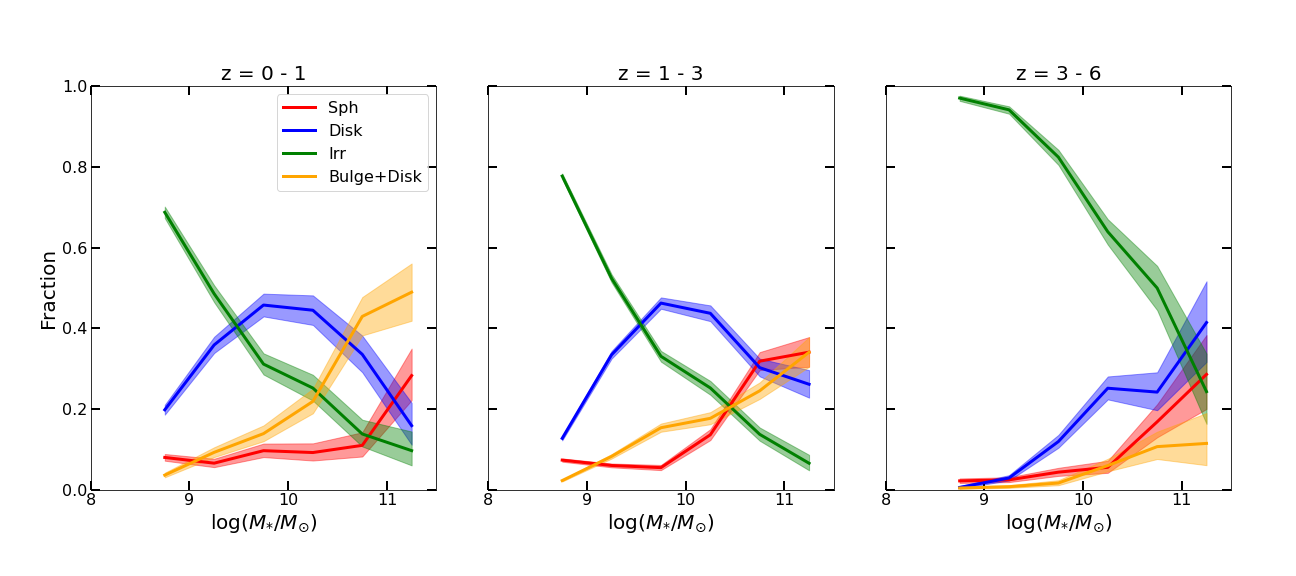}}
\vspace{-1.cm}\vspace{-1.cm}
\subfigure{\includegraphics[width=\linewidth]{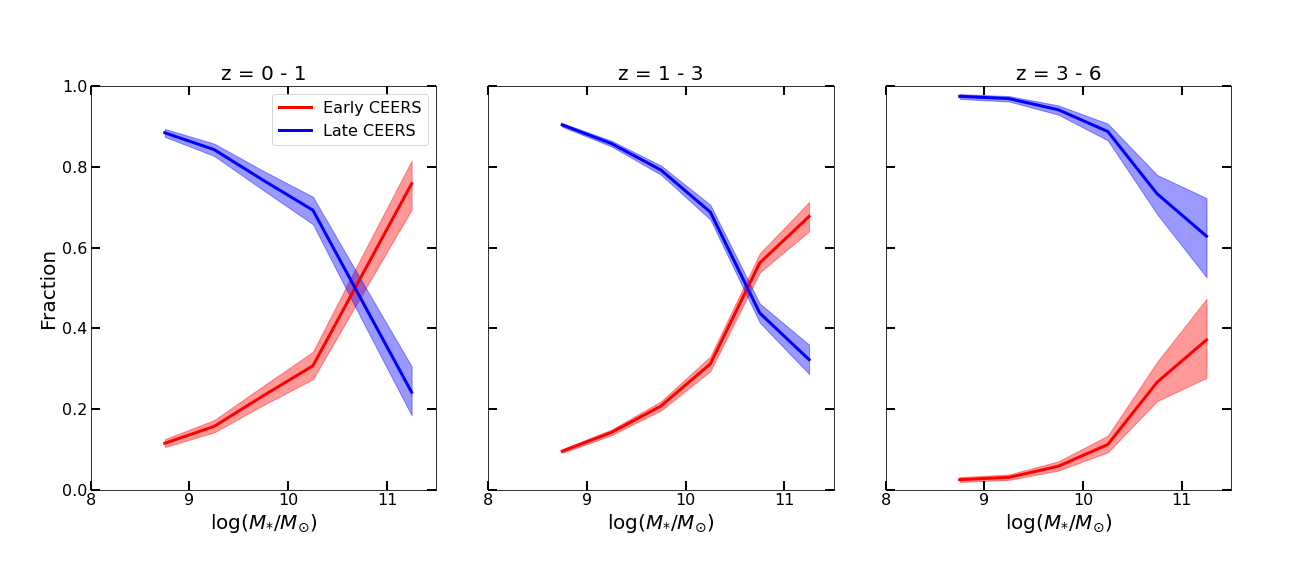}}\vspace{-1cm}\vspace{-0.7cm}
\subfigure{\includegraphics[width=\linewidth]{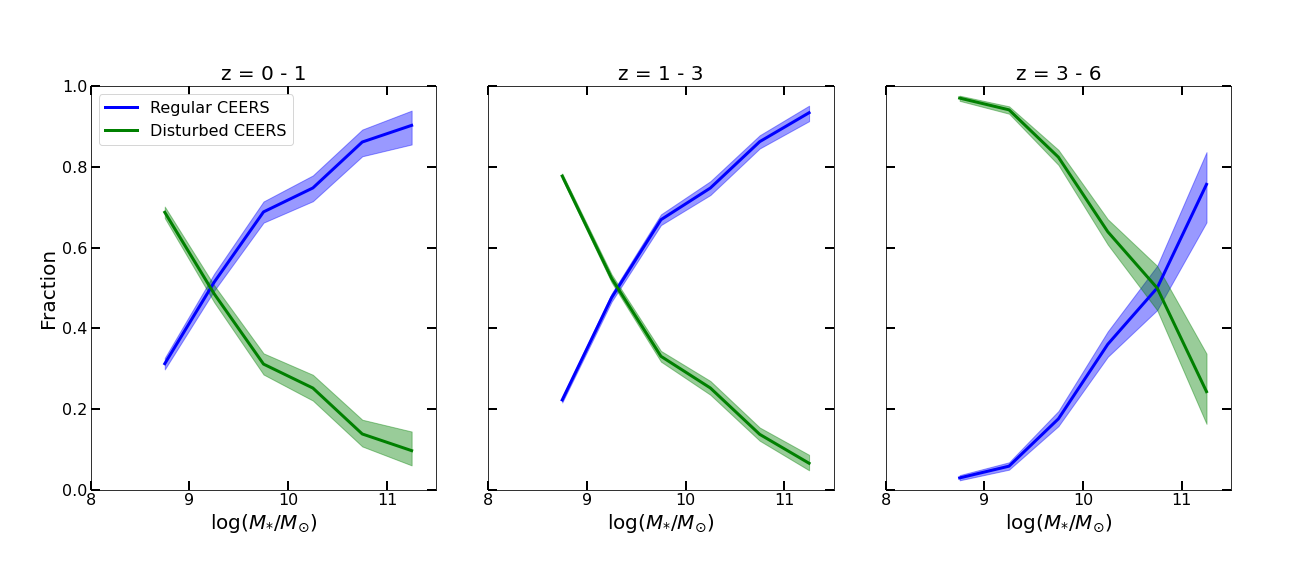}}
    \caption{Evolution of the fractions of different morphological types in rest-frame $\sim0.8-1\mu m$ as a function of stellar mass and redshift. Each panel shows a redshift bin as labeled. Filters F200W, F356W and F444W are used to infer galaxy morphology in the redshift bins $0<z<1$, $1<z<3$ and $3<z<6$ respectively. Top row: Fractions in 4 morphological classes: spheroids (red), disks (blue), bulge+disk (orange) and peculiar or irregular (green). Middle row: Fractions in two broad classes: disk dominated (blue) and bulge dominated (red). Bottom row: Fractions in two broad classes: regular (blue) and disturbed (green).  } 
    \label{fig:morph_fractions_rest_frame}
\end{figure*}

\begin{figure*}
\centering

\subfigure{\includegraphics[width=\linewidth]{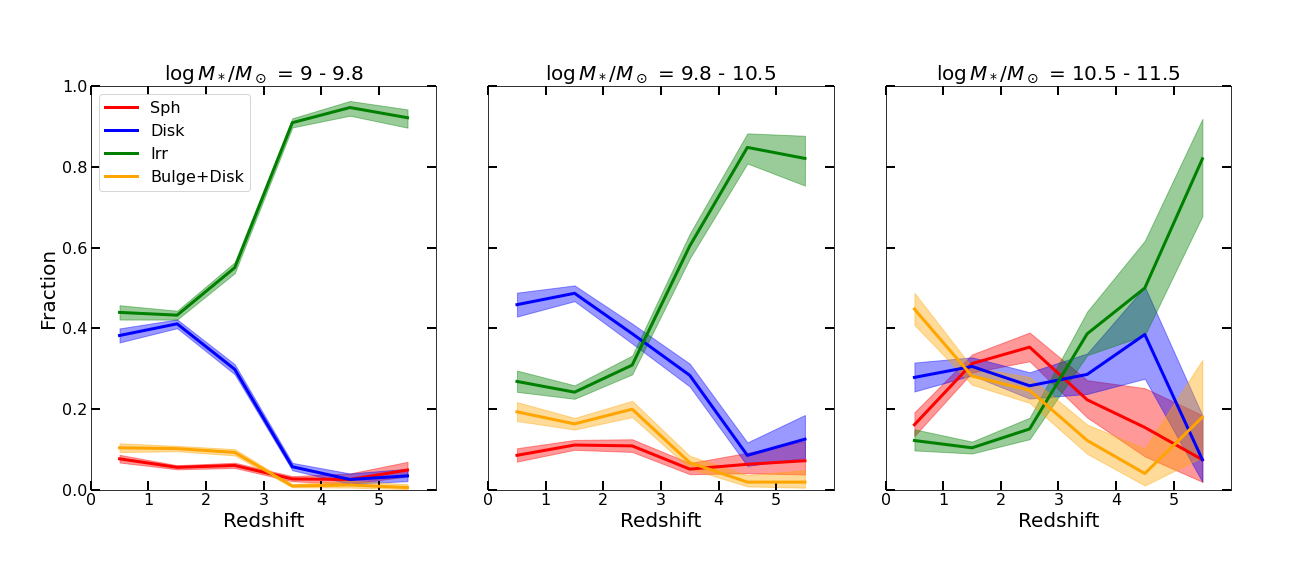}}
\vspace{-1.cm}\vspace{-1.cm}
\subfigure{\includegraphics[width=\linewidth]{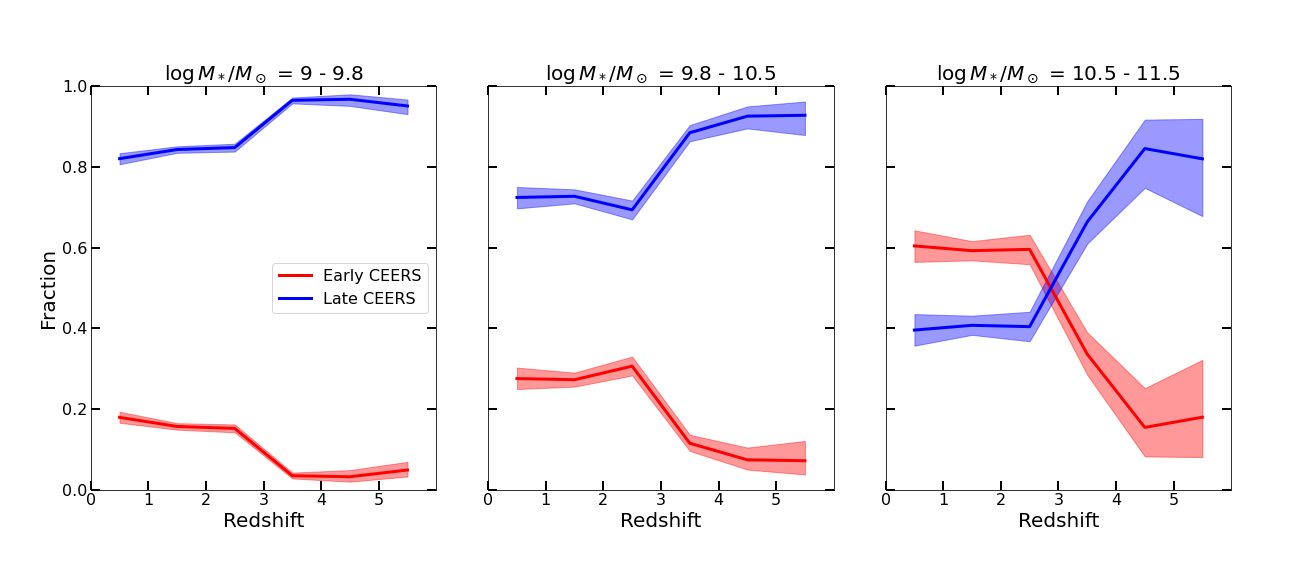}}\vspace{-1cm}\vspace{-0.7cm}
\subfigure{\includegraphics[width=\linewidth]{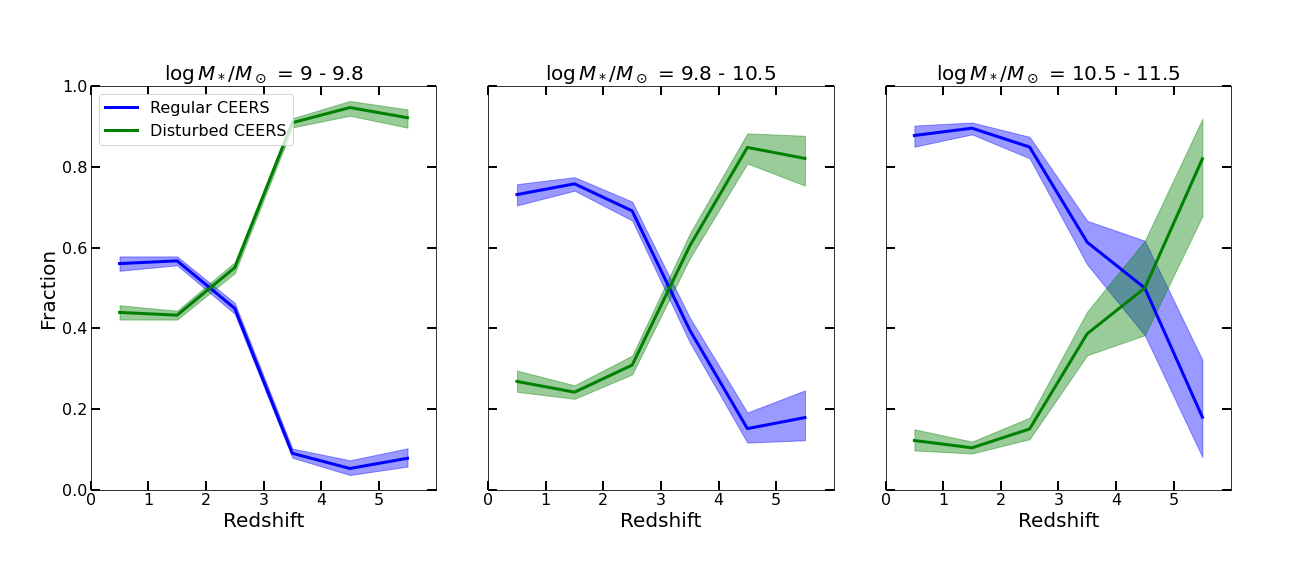}}
\caption{Evolution of the fractions of different morphological types in rest-frame $\sim0.8-1\mu m$ as a function of stellar mass and redshift. Filters F200W, F356W and F444W are used to infer galaxy morphology in the redshift bins $0<z<1$, $1<z<3$ and $3<z<6$ respectively. Each panel shows a stellar mass bin as labeled. Top row: Fractions in 4 morphological classes: spheroids (red), disks (blue), bulge+disk (orange) and peculiar or irregular (green). Middle row: Fractions in two broad classes: disk dominated (blue) and bulge dominated (red). Bottom row: Fractions in two broad classes: regular (blue) and disturbed (green).  } 
\label{fig:morph_fractions_z_rest_frame}
\end{figure*}

We examine how the fractions of different morphological types depend on stellar mass and redshift using the new JWST-based morphologies derived in this work (see Figures~\ref{fig:morph_fractions_rest_frame} and~\ref{fig:morph_fractions_z_rest_frame}). To define the morphologies, we use the $F200W$ filter at $z<1$, $F356W$ at $1<z<3$, and $F444W$ at $3<z<6$, approximately in the rest-frame $\sim0.8-1\mu m$ (Figure~\ref{fig:rest_wl}). This should be mostly sensitive to the emission of old stars and therefore more closely linked to stellar mass. The price to pay - as previously mentioned - is that the spatial resolution is worse in the redder filters potentially biasing the measured evolution of morphologies. Nevertheless, as shown in Figure~\ref{fig:rest_wl}, the resolution remains reasonably constant around $\sim0.5$ kpc from $z\sim1$ to $z\sim6$ and only goes down to $\sim0.2$ kpc at $z<1$. 

We observe some well-known trends. The fraction of bulge-dominated galaxies (early-type) shows a strong correlation with stellar mass, with the number densities of early-type galaxies steadily increasing above $\sim10^{10.3}$ solar masses. The behavior is surprisingly similar at all redshifts probed, suggesting similar physical processes for bulge formation at all epochs.  Also, the fact that the trends change smoothly with redshift, is an indication that there are no significant biases induced by the use of different filters with different spatial resolutions. The main difference is a change in normalization, with early-type galaxies becoming more abundant at fixed stellar mass at later times. At $\sim10^{11}$ solar masses, $\sim40-50\%$ of galaxies are bulge-dominated at $z>3$, while the fraction increases to $\sim70\%$ at $z<1$ for the same stellar mass. Figure~\ref{fig:morph_fractions_z_rest_frame} shows in fact that early-type galaxies start dominating the massive end of the galaxy population from $z\sim3$. These results qualitatively agree with the recent findings by \citet{2021ApJ...913..125C}, who reported a bimodality in bulge formation, with a first wave of bulges ($\sim30\%$) already in place at $z\sim6$.~\cite{2022arXiv221001110F} tend to find a smaller fraction of bulge dominated galaxies at all redshifts using a similar dataset but based on pure visual classifications. It might be because they did not plot the most massive galaxies separately as we do here. The lower mass bins in Figure~\ref{fig:morph_fractions_z_rest_frame} show indeed that late-type galaxies dominate at all redshifts. 

Regarding the fraction of peculiar galaxies, we also measure very similar trends at all redshifts. The abundance of irregular galaxies is a strong function of stellar mass at all redshifts, with low-mass galaxies being predominantly peculiar. The stellar mass threshold below which the galaxy population starts to be dominated by irregular galaxies decreases with time. Galaxies less massive than $\sim10^{10.5}$ solar masses are irregular at $z>3$, while this is true only for galaxies less massive than $\sim10^{9}$ at $z<1$.  It is interesting to see that the abundance of irregular galaxies is still measured to increase with resdshift even when probing the rest-frame NIR. This suggests that, at early epochs, the distribution of stellar mass is also perturbed and that it is not only a consequence of the presence of bright star-forming regions emitting in the UV. However, as discussed in Section~\ref{sec:HST_JWST}, the classification of a galaxy as irregular or peculiar does not only depend on wavelength but is also rather noise-sensitive. The result might therefore be biased because galaxies at fixed stellar mass appear fainter at high redshift. It is also worth noticing, though, that the impact of the SNR on the reported fractions is generally small (see Figure~\ref{fig:frac_CEERS_CANDELS}) because it affects essentially large galaxies that represent a small fraction of the galaxy population. Our results qualitatively agree with the trends reported by~\cite{2022arXiv221001110F} in the sense that \emph{regular} galaxies tend to dominate at lower redshifts, but they do not find such clear trends. In particular, they find that low mass galaxies ($\log M_*/M_\odot<9$) are predominantly disks while our classification tends to classify the majority of low mass galaxies as irregulars at all redshifts (bottom left panel of Figure~\ref{fig:morph_fractions_z_rest_frame}). The differences might be a reflection of the somehow loose definition of the peculiar class. A one-to-one comparison between the two classifications should provide more insights on the origins of the differences.  

\subsection{Morphology-Quenching relation}

In this section, we investigate the evolution of the relationship between galaxy morphology and star-formation activity, using the new NIR rest-frame morphologies based on JWST, from $z\sim6$.

\subsubsection{Definition of Quenched and Star-Forming Samples}

\begin{figure*}
    
{\includegraphics[width=0.33\linewidth]{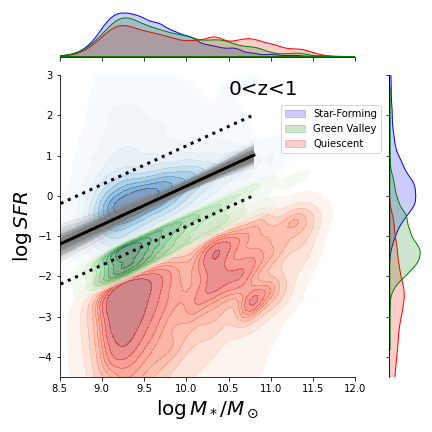}}
{\includegraphics[width=0.33\linewidth]{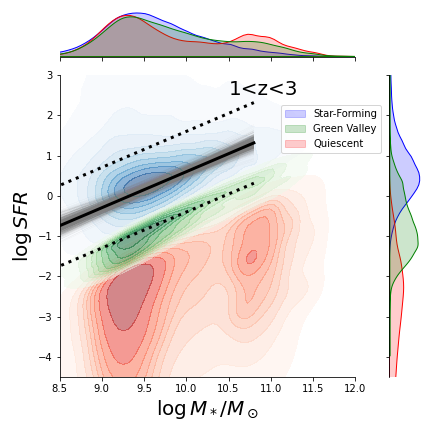}} 
{\includegraphics[width=0.33\linewidth]{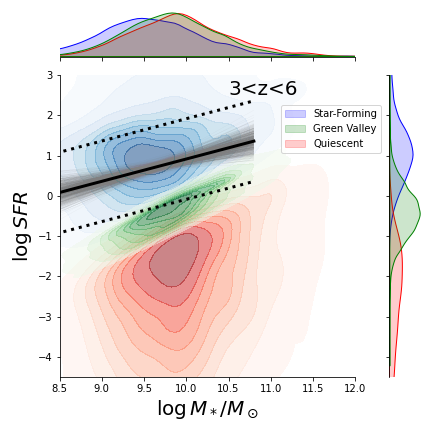}} 
{\includegraphics[width=0.33\linewidth]{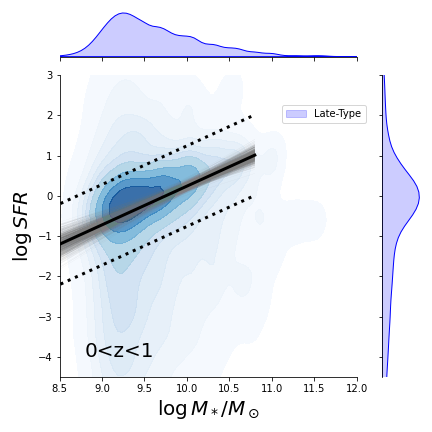}}
{\includegraphics[width=0.33\linewidth]{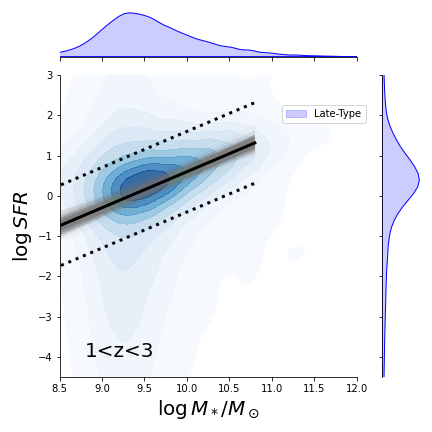}} 
{\includegraphics[width=0.33\linewidth]{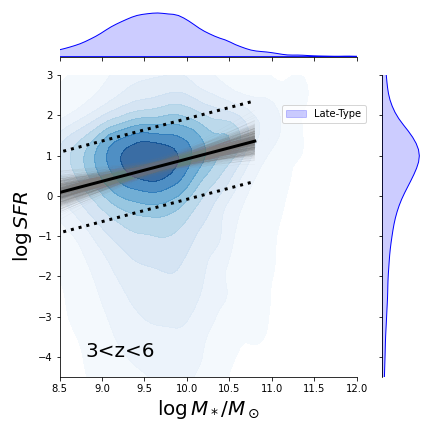}} 

{\includegraphics[width=0.33\linewidth]{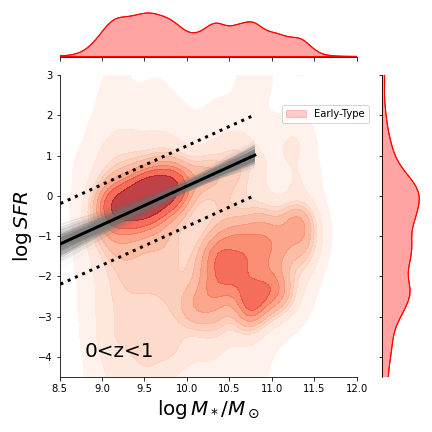}}
{\includegraphics[width=0.33\linewidth]{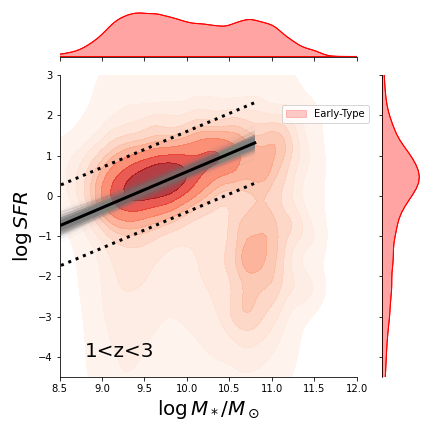}} 
{\includegraphics[width=0.33\linewidth]{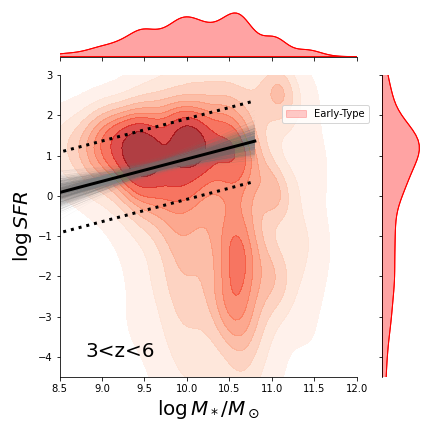}}

    \caption{Stellar mass - star formation rate ($M_\odot.yr^{-1}$) plane in different redshift bins as labeled. The top row shows the whole sample analyzed in this work divided in star-forming (blue contours), green valley (green contours) and quiescent galaxies (red contours) based on the mass doubling time (see text for details). The middle row shows the distribution of late-type galaxies and the bottom row the one for early-type galaxies. In all panels, the solid line shows the mean of the posterior power-law fit to star-forming galaxies and the gray shaded regions indicates random samples of the posterior distribution. The dotted lines indicate the location $\pm1$ dex around the main sequence.} 
    \label{fig:SFR_mstar}
\end{figure*}

To begin, we define quiescent and star-forming galaxies, following the definition of~\cite{2022ApJ...926..134T}, which is based on the timescale for doubling the stellar mass, given the instantaneous star formation rate at the time of observation.\cite{2022ApJ...926..134T} measure this timescale by defining the mass-doubling number $D(z)=$sSFR(z)$\times t_H(z)$, where sSFR(z) and $t_H(z)$ represent the specific star-formation rate and the Hubble time at the time of observation, respectively. Therefore, $D(z)$ measures the number of times the stellar mass doubles within the age of the universe, at a constant sSFR. To account for uncertainties in both the measurements of star formation rates and stellar masses, we sample 100 times from the posterior distributions of both quantities estimated by \textsc{Dense Basis} . For simplicity, we assume a Gaussian posterior distribution with standard deviation equal to half the difference between the 84th and 16th quantiles, as estimated by \textsc{Dense Basis} (see section~\ref{sec:data} for more details). The top row of Figure~\ref{fig:SFR_mstar} displays the contours of the distribution of the 100 samples in the $\log M_*-\log$ SFR plane, for different redshift bins. The SFR is averaged over a timescale of $100$ Myrs, where star-forming galaxies are defined as those with $D(z)>1/3$, transitionning galaxies those with $1/20<D(z)<1/3$ and quiescent galaxies those with $D(z)<1/20$, following~\cite{2022ApJ...926..134T}.

Next, we fit the distribution of star-forming galaxies ($D(z)>1/3$) using a power law with two parameters $\alpha$ and $\beta$:

$$ \log SFR = \alpha\times(\log M_*/M_\odot-10.5) +\beta$$

We estimate the posterior distributions $p(\alpha,\beta | \{\log M_*,\log SFR\})$ in three redshift bins - $0<z<1$, $1<z<3$, and $3<z<6$ - using an amortized likelihood-free inference approach with a Masked Autoregressive Neural Flow. We simulate $100,000$ samples, using a flat prior for both $\alpha$ and $\beta$, and using a Gaussian distribution for both $\log$ SFR and $\log M_*$. We then train a Neural Flow on the simulations to estimate $p$. We train three different flows for the three different redshift bins. To account for systematic uncertainties due to the neural network architecture, we train five additional random variations of the masked autoregressive flow in each redshift bin. We find that all models provide consistent results and choose one of them as the primary estimator. The mean and 16th and 84th quantiles of $\alpha$ and $\beta$ are reported in Table~\ref{tbl:fits}.

\begin{table}[h]
    \centering
    \caption{Slope ($\alpha$) and normalization ($\beta$) of the power-law fit 
    to the star-forming main sequence in different redshift bins}
    \begin{tabular}{|c|c|c|c|}
        \hline
         & \makecell{$0<z<1$} & \makecell{$1<z<3$} & \makecell{$3<z<6$} \\
         \hline
         &  &  &  \\
        $\alpha$ & \makecell{$0.96^{+0.11}_{-0.12}$} & \makecell{$0.89^{+0.07}_{-0.07}$} & \makecell{$0.55^{+0.16}_{-0.15}$} \\
         &  &  &  \\
        $\beta$ & \makecell{$0.72^{+0.09}_{-0.08}$} & \makecell{$1.04^{+0.09}_{-0.10}$} & \makecell{$1.19^{+0.15}_{-0.15}$} \\
         &  &  &  \\
        \hline
    \end{tabular}
    \label{tbl:fits}
\end{table}

We observe an increase in the zeropoint ($\beta$) as we move to higher redshifts, consistent with previous studies (e.g., ~\citealp{2012ApJ...754L..29W}). Furthermore, we note a slight decrease in the slope of the star-forming main sequence in the highest redshift bin which is also within the ballpark of the many different values reported in the literature (see~\cite{2023arXiv230316234M} for a recent compilation). Figure~\ref{fig:SFR_mstar} shows that this decrease in slope could be driven by a larger scatter in star formation at fixed stellar mass and the lack of massive objects at these redshifts. Further investigation is necessary to fully understand this trend. However, our primary goal is to obtain a reasonable guess of the location of the main sequence, rather than to analyze the slope and normalization in detail, which is beyond the scope of this work.

The middle and bottom rows of figure~\ref{fig:SFR_mstar} display the distribution of late-type and early-type galaxies in the $\log M_*-\log SFR$ plane. Consistent with expectations, the vast majority of disk-dominated galaxies are located within 1 dex of the star-forming main sequence. In contrast, early-type galaxies are located in two distinct regions. At the high-mass end, they tend to lie below the main sequence in the region where quenched galaxies are found, confirming that massive quenched galaxies tend to be bulge-dominated. Interestingly, this trend seems to hold even in the highest redshift bin. Additionally, the bottom row of figure~\ref{fig:SFR_mstar} reveals a significant population of low-mass ($\log M_*/M_\odot<10$) star-forming bulge-dominated galaxies located at all redshifts, which we investigate further in the following subsections.

\subsubsection{Fraction of early-type galaxies}

\begin{figure*}
    
\includegraphics[width=\linewidth]{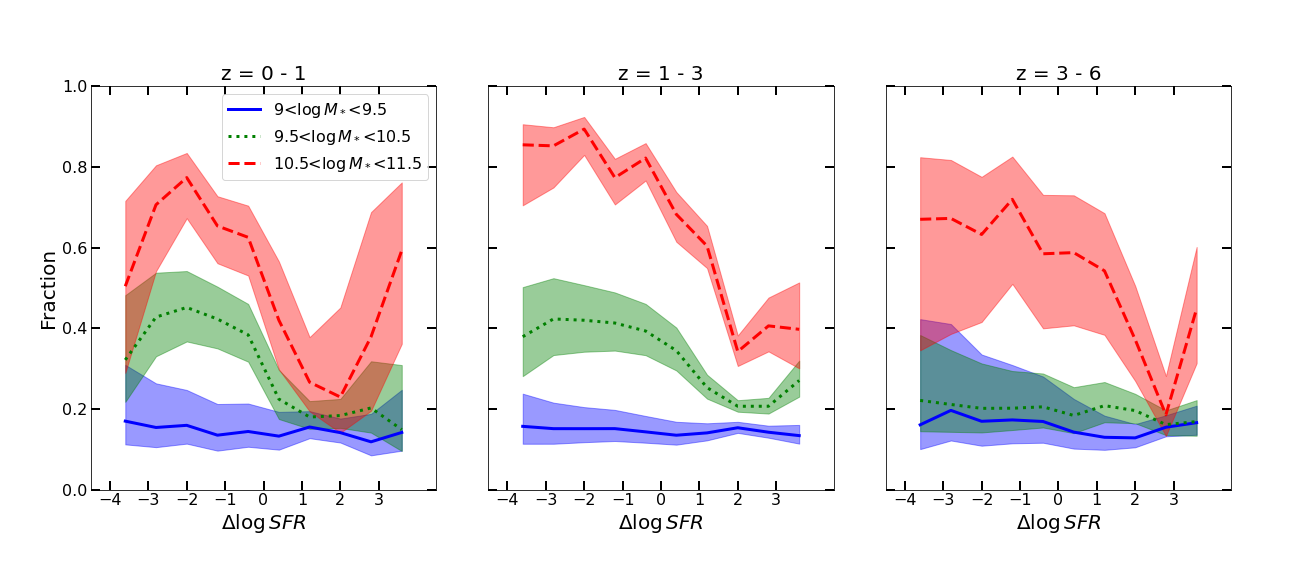}
\caption{Fraction of early-type galaxies as a function of $\Delta\log$ SFR. Each panel shows a different redshift bin - $0<z<1$,$1<z<3$,$3<z<6$ - from left to right. The red dashed, green dotted and solid blue lines show galaxies with stellar masses of $10.5<\log M_*/M_\odot<11.5$, $9.5<\log M_*/M_\odot<10.5$, $9.5<\log M_*/M_\odot<10.5$ respectively. } 
    \label{fig:ETF_deltaSFR}
\end{figure*}

Using the power-law fits discussed in the previous subsection, we compute the fraction of early-type galaxies as a function of the distance to the main sequence, $\Delta\log$ SFR, in different stellar mass bins, and present the results in Figure~\ref{fig:ETF_deltaSFR}. Here, $\Delta\log$ SFR is calculated as the difference between a galaxy's SFR and the location of the main sequence corresponding to its stellar mass. We find that the fraction of bulge-dominated systems among massive galaxies ($\log M_*/M_\odot>10.5$) increases as we move below the main sequence at all redshifts. Despite the large uncertainties due to small statistics, we observe that $\sim70\%-90\%$ of massive galaxies 2-3 dex below the main sequence are early-type. These results suggest that the morphology-quenching relation is already in place for the most massive galaxies at $z\sim5$. At intermediate stellar masses, we also observe a moderate increase in the early-type fraction below the main sequence, at least up to $z\sim3$. However, at $z>3$, the fraction remains consistently low ($<10\%$) regardless of star formation activity, indicating that morphological transformations have not yet occurred in significant numbers by $z\sim5$. We do not observe any dependence of the early-type fraction on star formation activity for low mass galaxies within the probed redshift range.

\subsubsection{Star-forming early-type galaxies}

We analyze the distribution of the star-formation rate at fixed morphological type. Although the fraction of low-mass early-type galaxies is small at low and intermediate masses, Figure~\ref{fig:SFR_mstar} shows that the majority of those are star-forming, in contrast to what happens at the high-mass end. 

\begin{figure*}
{\includegraphics[width=0.50\linewidth]{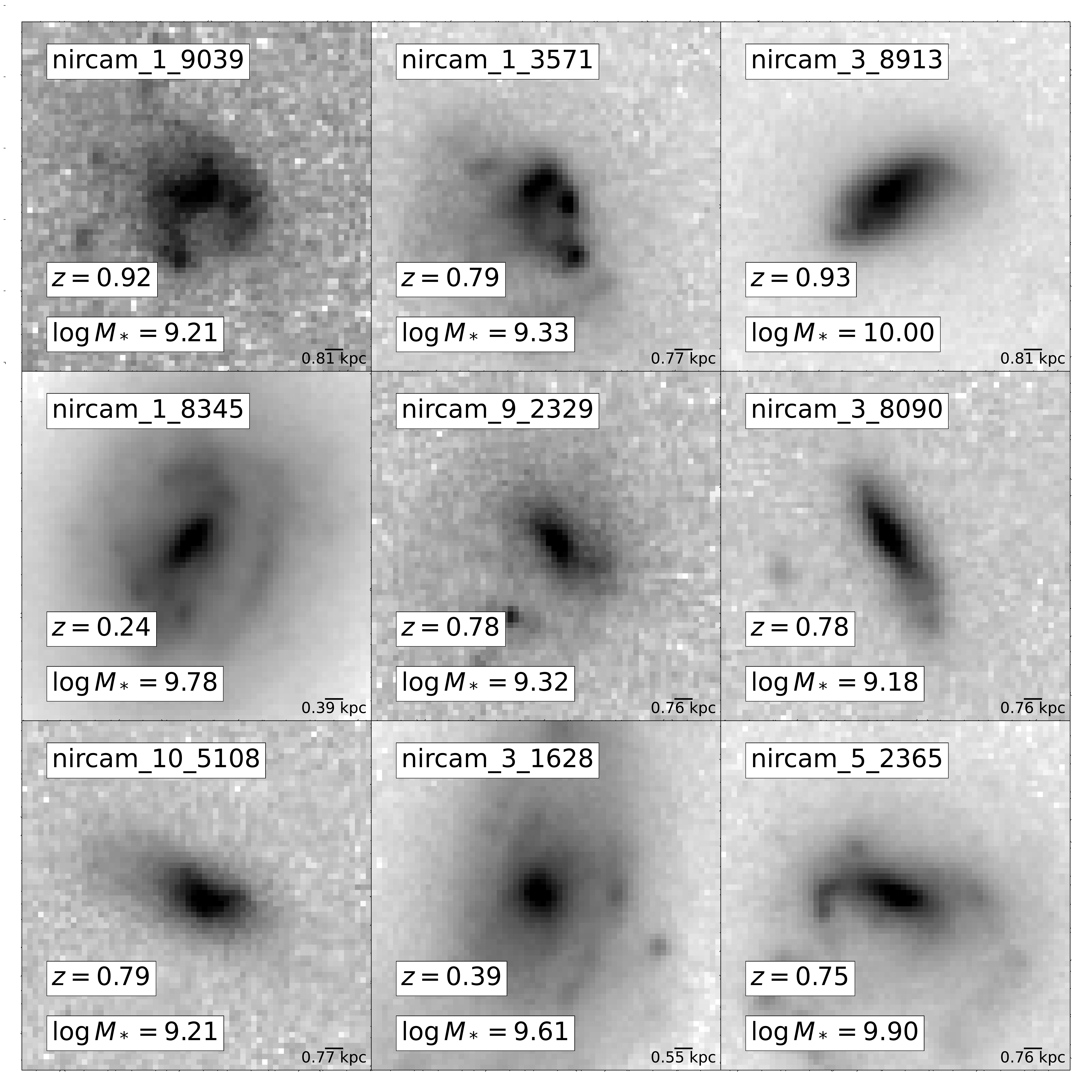}}
{\includegraphics[width=0.50\linewidth]{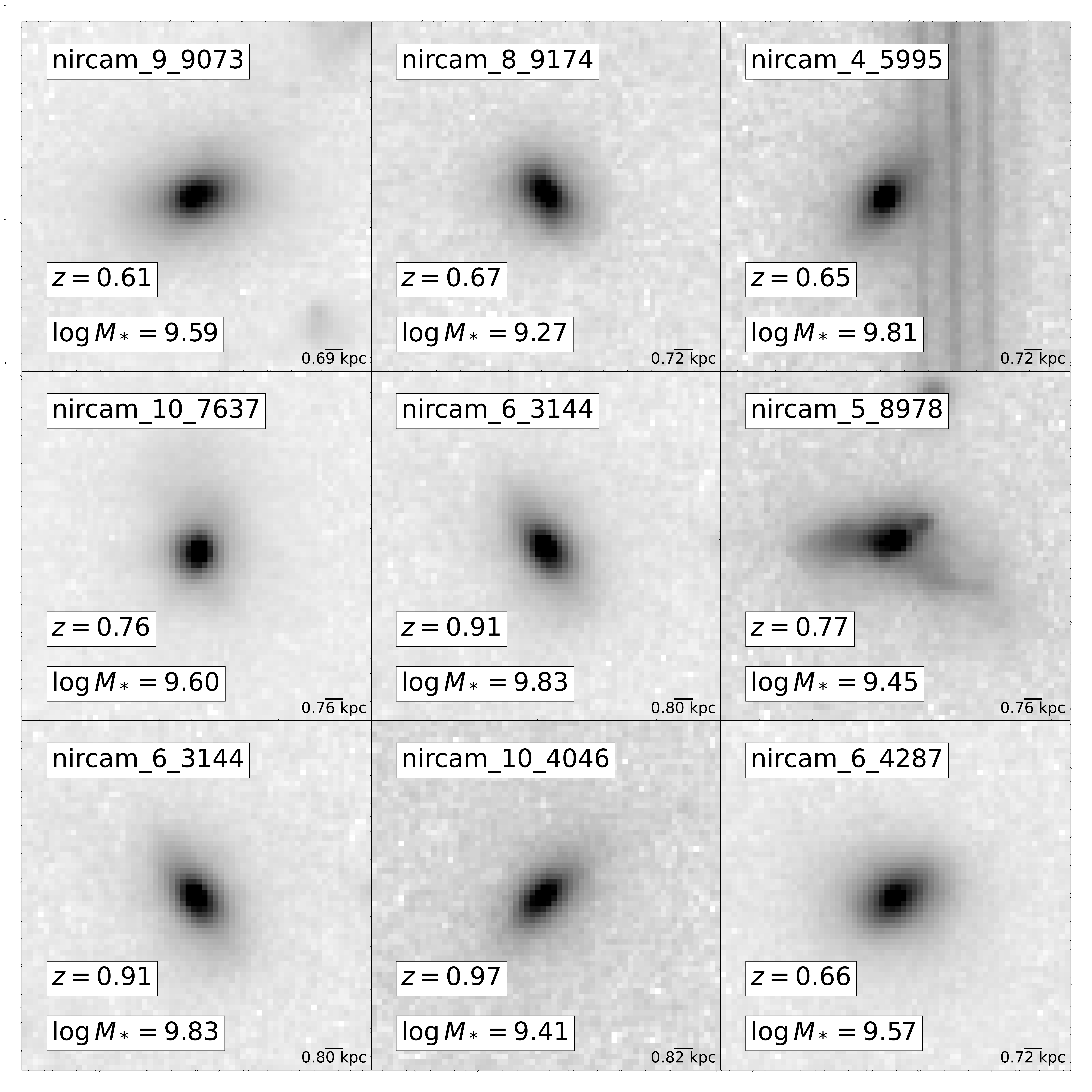}} 
       \caption{Random example stamps of low mass ($9<\log M_*/M_\odot<10$) star-forming galaxies with late-type (left panel) and early-type morphologies (right panel). Images are in the $F200W$ filter.} 
    \label{fig:SF_ETGs}
\end{figure*}

To investigate further the nature of these objects and check whether this is a consequence of classification errors, we show in Figure~\ref{fig:SF_ETGs} example stamps of late-type and early-type star-forming galaxies ($D(z)>0.33$) with similar stellar mass. We only show examples of $z<1$ galaxies to better appreciate the morphological differences. The figure clearly shows distinct morphologies for both populations. As expected, late-type galaxies are more extended and present more structure than early-type galaxies, whose light profile is smoother and more compact. An inspection of the half-light radii and S\'{e}rsic indices distributions of both populations also reveals that early-type galaxies are, on average, more compact and have higher S\'{e}rsic indices. Both results confirm the distinct morphological nature of these star-forming galaxies. We speculate that the population of low-mass early-type galaxies might be experiencing a phase of central star-formation activity (\emph{blue nugget}) as reported in several previous works (e.g.,~\citealp{2023arXiv230212234L,2018ApJ...858..114H, 2016ApJ...827L..32B}, and references therein). However, an analysis of the non-parametric SFHs inferred by \textsc{Dense Basis} does not show any clear differences between the two populations given the large uncertainties at the current S/N. Therefore we cannot firmly conclude that they have clear different formation pathways. Future deep observations targeting these high-redshift galaxies might enable us to break these degeneracies and better constrain their SFHs. 

\section{Discussion}

\label{sec:disc}

\subsection{The differences between HST and JWST-based galaxy morphologies}
\label{sec:disc_HST_JWST}

Galaxy morphology is a powerful proxy for physical processes in galaxies, but it is also well known to be significantly affected by multiple observational effects such as signal-to-noise ratio (SNR), spatial resolution, and cosmological dimming. Most results published in the past decades have been based on HST data, which, despite being a powerful space-based telescope, has been pushed to its limits for analyzing galaxy structures beyond the local universe. In particular, many works based on HST data (e.g., \citealp{2013MNRAS.428.1460B,2014ApJ...788...11L,2014ApJ...788...28V,2016MNRAS.462.4495H}) have shown that massive bulge-dominated galaxies exist at least from $z\sim3$, suggesting that effective dissipative processes efficiently form bulges and destroy disks in the early universe (e.g., \citealp{2007ApJ...658..710N,2016MNRAS.457.2790T, 2023arXiv230212234L}). However, these results may have been partially biased by the depth of observations coupled with cosmological dimming, which makes it difficult to detect a disk component. Our work, based on JWST data with significantly better sensitivity and a factor of $\sim2$ increased spatial resolution, tends to confirm the morphological classifications performed with HST for the vast majority of objects. As expected, we find that JWST classifications tend to present more disks, but the differences affect only a small number of objects that do not change the main trends.

Another key result of the past decades has been that galaxy morphologies become more irregular and disturbed as we move to higher redshifts (e.g., \citealp{1996MNRAS.279L..47A,2000ApJ...529..886C,2015ApJ...800...39G,2016MNRAS.462.4495H}). However, there has been some debate about whether this is due to more disturbed kinematics of stars or if it is biased because HST filters probe bluer light at high redshift. A disturbed appearance of the light distribution can also be caused by poor SNR, which tends to create noise fluctuations that can be interpreted as irregularities in the surface brightness distribution. Kinematic studies of ionized gas at high redshift (e.g., \citealp{2008ApJ...687...59G,2019ApJ...886..124W}) have indeed suggested that the majority of galaxies at $z\sim1-2$ present clear rotation patterns, suggesting a more regular kinematic structure than what is inferred from the UV rest-frame light, although with higher velocity dispersion than local disks (e.g., \citealp{2016ApJ...830...14S}). Some works have tried to infer the stellar mass distribution from resolved pixel-based SED fitting (e.g., \citealp{2012ApJ...753..114W}), finding a smoother distribution than light. Other works have also estimated that the contribution of bright UV clumps to the stellar mass is rather modest~\citep{2020MNRAS.499..814H}.

JWST data, with its improved sensitivity, spatial resolution, and longer wavelength coverage, allows us to revisit the issue of quantifying the abundance of galaxies with irregular morphologies using the rest-frame NIR, which more closely tracks stellar mass. Our work confirms that resolution and SNR play an important role in defining the class of irregular galaxies. When compared consistently, JWST-based classifications tend to find up to $\sim30\%$ fewer irregular galaxies than HST-based ones on the same objects. However, despite this decrease, the reported fractions as a function of redshift look very similar between HST and JWST - when compared at similar rest-frame wavelengths - because it only affects large objects with low SNR per pixel which represent a small fraction of the galaxy population. The fraction of irregular galaxies is hence confirmed to increase and does not seem to be purely driven by an observational bias at least at first order.

As previously mentioned, JWST does not only enable the quantification of the effect of SNR and resolution in the classification of irregular galaxies, but also allows us to look at morphologies in longer wavelengths than HST and therefore quantify how much of the increase in the fraction of irregular galaxies is driven by the light emitted by young stars. First works using JWST data have suggested that the fraction of regular disks might have been underestimated by HST \citep[e.g.,][]{2022ApJ...938L...2F,2022arXiv221014713K, 2023ApJ...942L..42R}. We confirm in this work that $\sim30\%$ of the galaxies tend to appear less disturbed when looked in the NIR rest-frame. This would point towards a smoother distribution of mass than previously measured with HST and more in agreement with pixel-based SED fitting results \citep{2012ApJ...753..114W}. We find that undisturbed disk-like morphologies are rather common at the high stellar mass end up to $z\sim5$. However, low mass galaxies are still found to be irregular in their vast majority even in the NIR rest-frame, suggesting more perturbed kinematics than in the local universe. This is in slight disagreement with the measurements of~\cite{2022arXiv221001110F} who find that disks still dominate at the low mass end. The discrepancies might be a consequence of the different definitions of peculiar galaxies. Future spectroscopic JWST-based observations should allow us to further constrain the physical properties of these galaxies.


\subsection{The onset of the formation of disks and bulges}
\label{sec:disc_onset}

Understanding when and how bulges form remains a fundamental question in the field of galaxy formation for which new JWST is expected to provide new insight. Our work indeed provides a new look into the emergence of morphological diversity from $z\sim6$ as seen by JWST at longer wavelengths than ever before with HST.

\begin{figure*}
\centering
\subfigure{\includegraphics[width=\linewidth]{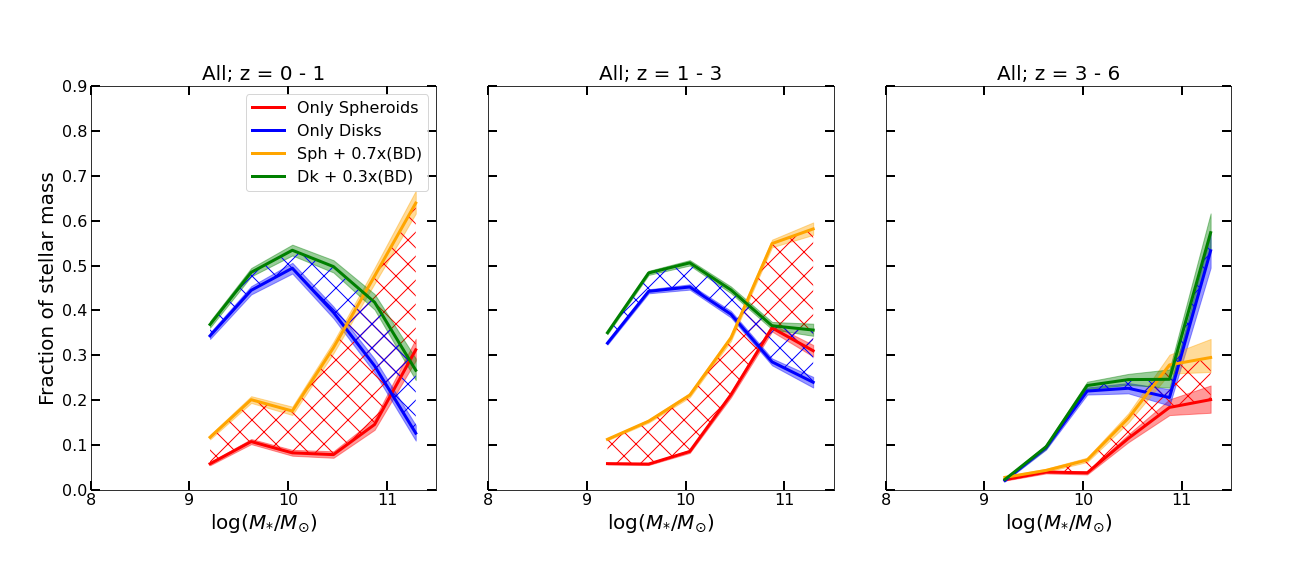}}
\vspace{-1.cm}\vspace{-1.cm}
\subfigure{\includegraphics[width=\linewidth]{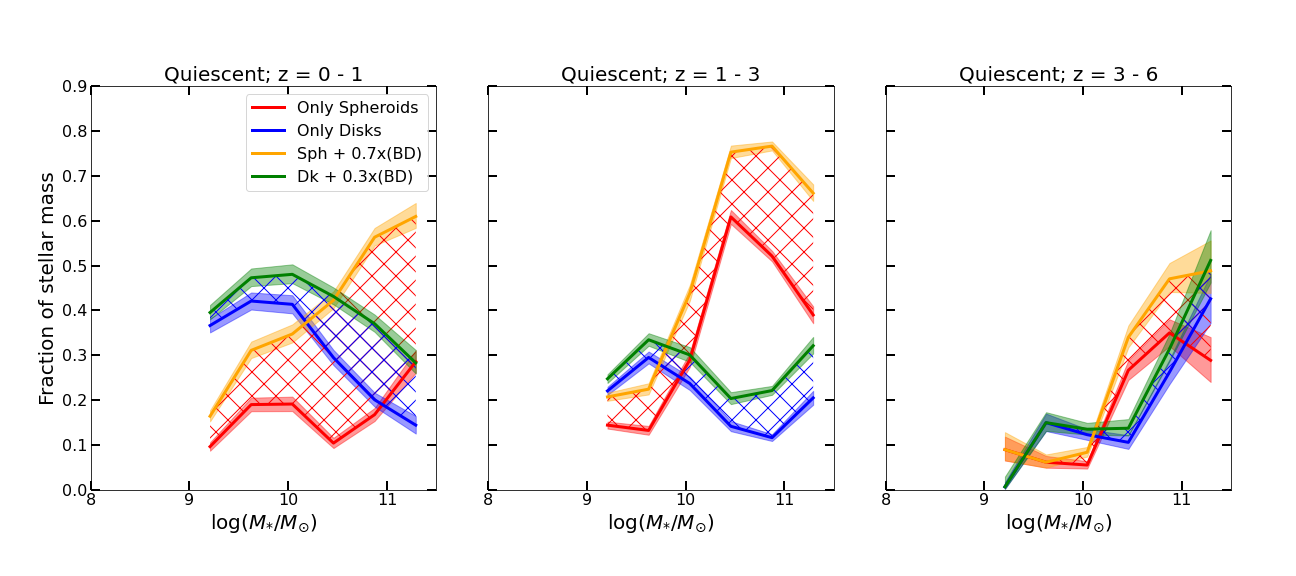}}
\vspace{-1.cm}\vspace{-1.cm}
\subfigure{\includegraphics[width=\linewidth]{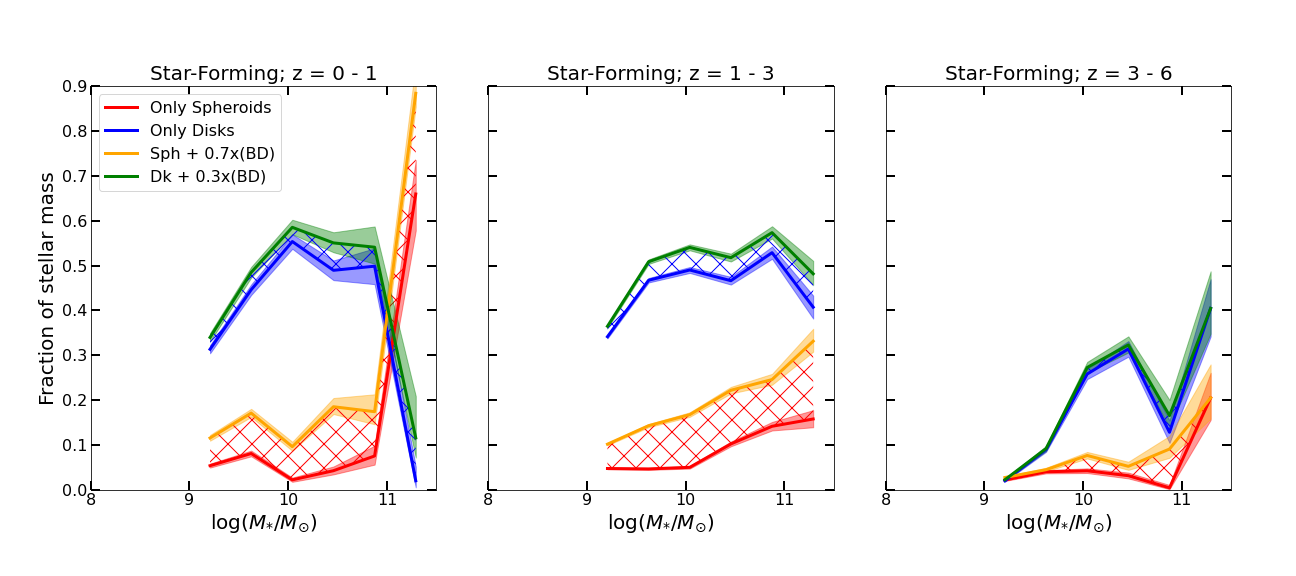}} \vspace{-1cm}
    \caption{Fraction of stellar mass in bulges (red hashed region) and disks (blue hashed region) for all (top row), quiescent (middle row) and star-forming (bottom row) galaxies in different redshift bins as labeled. } 
    \label{fig:mass_bulges_disks}
\end{figure*}

Numerous previous works based on HST have shown that, even if the fraction of bulge-dominated galaxies among massive galaxies decreases as we move at high redshift, massive bulge-dominated galaxies are found at least from $z\sim3$, suggesting that bulge formation starts very early (e.g., ~\citealp{2021ApJ...913..125C}). We extend the census of bulges up to $z\sim6$. We confirm a decreasing trend of their abundance. At $z>3$, the majority of massive galaxies are not bulge-dominated. However, we still measure $\sim20\%$ of bulge-dominated galaxies at $z\sim6$ suggesting that massive galaxy formation is well evolved already at these early epochs. Our work also shows that the connection between morphology and quenching, which has been reported in a large variety of works up to $z\sim3$ (e.g.,~\citealp{2008ApJ...677L...5V,2012ApJ...745..179W,2014ApJ...788...28V,2014MNRAS.441..599B,2017ApJ...840...47B} and many others), is already in place at $z\sim5$. The fraction of bulge-dominated galaxies below the main sequence is close to $\sim60\%$ also pointing towards a rapid assembly of stellar mass and efficient feedback mechanisms to quench star formation a few hundred million years after the Big Bang. In Figure~\ref{fig:mass_bulges_disks}, we make an attempt to estimate the fraction of the stellar mass in bulges and disk components as a function of stellar mass and redshift. We do so by assuming that, for galaxies classified as spheroids, all the mass is in the form of bulges, while for bulge+disk systems, we set it to an arbitrary number of $0.7$ of the total mass of the galaxy. We estimate the amount of mass in bulges lies between the lower limit estimate from only spheroids and the upper limit, which comes from the sum of spheroids and composite systems. We understand that this is a very first order approximation with strong assumptions, but we think can be used at least to explore the main trends. We see that up to $z\sim3$, $\sim60\%$ of the stellar mass in quiescent galaxies more massive than $10^{10}$ solar masses is in bulges, which is compatible with previous HST-based measurements (e.g.,~\citealp{2014ApJ...788...11L,2016MNRAS.462.4495H}). At $z\sim5$, the fraction decreases to $\sim30\%$, which suggests we are approaching the onset of massive bulge formation. However, there is still a significant population of massive bulges in place at these early epochs. A careful comparison with the predictions of numerical simulations can help put some additional constraints on feedback models in the early universe (e.g.,~\citealp{2023arXiv230304827D}).

Interestingly, for star-forming galaxies, the fraction of stellar mass in bulges remains of the order of $\sim10\%$ at all epochs, also confirming previous findings~(e.g.,\citealp{2022MNRAS.513..256D}). Although the abundance of these star-forming bulges, sometimes referred as \emph{blue nuggets}, is small, numerous works have suggested they can represent an important phase for bulge build-up (see~\citealp{2023arXiv230212234L} for a nice review on the topic). Our work confirms the presence of these systems. Although there might still be a disk component which is not detected, the fact that we see this star-forming bulge dominated population at low redshift and with deeper imaging than with HST, suggests that that population exists. Whether the morphological transformations and quenching are causally connected (e.g.,\citealp{2016MNRAS.457.2790T,2018ApJ...853..131L,2020ApJ...897..102C,2022MNRAS.513..256D,2022ApJ...929..121C,2023arXiv230212234L}) or whether it is just a consequence of progenitor bias (e.g.,\citealp{2016ApJ...833....1L}) is not something that our results can directly address. However, the fact that a galaxy bimodality already exists at these very early epochs, where the impact of progenitor bias should be rather limited given the short amount of elapsed time, suggests that some degree of morphological transformation is taking place before or after quenching.

Another important question is when the first disks are formed. The current cosmological model predicts that massive haloes are assembled by the merging of smaller ones. Simulations show that low mass galaxies at high redshift have indeed perturbed kinematics or even prolate shapes (e.g.,\citealp{2016MNRAS.458.4477T}) which has also been hinted by some observational studies (e.g.,\citealp{2014ApJ...792L...6V,2019MNRAS.484.5170Z,2023arXiv230207277V}). ~\cite{2012ApJ...758..106K} and~\cite{2017ApJ...843...46S} also find that the abundance of rotationally supported systems increases with stellar mass with a transition mass increasing with redshift. A recent theoretical work by~\cite{2020MNRAS.493.4126D} suggests that gas disks only survive above a characteristic stellar mass of $\sim10^9$ since for lower mass systems the frequent mergers change the spin in less than an orbital time preventing disk formation. Although it is extremely difficult to infer the true disky nature of a system solely based on its apparent morphology (see e.g.,\citealp{2023arXiv230207277V} for a discussion on this), our results looking at the NIR rest-frame morphologies suggest that massive \emph{unperturbed} disk objects do exist at $z\sim5$ (see e.g.,~\citealp{2020Natur.584..201R,2021Sci...371..713L} for similar conclusions based on gas). We also find that the abundance of galaxies morphologically identified as disks strongly depends on stellar mass in qualitative agreement with theoretical predictions and gas kinematic studies up to $z\sim2$. Following an analogous procedure to what was done for the bulges, we quantify in Figure\ref{fig:mass_bulges_disks} the fraction of stellar mass in disks in star-forming galaxies as a function of stellar mass and redshift (irregular morphologies are not included in this estimate). Interestingly, we find that the fraction remains relatively constant across cosmic time between $\sim30\%$ and $\sim50\%$, suggesting that disk formation above a certain mass happens even at very early epochs. We emphasize that the measurements presented in this work do not allow us to firmly conclude on the true \emph{disky} nature of the galaxies classified as "disks". We recall that another potential source of bias could be related to SNR (see Subsection~\ref{sec:disc_HST_JWST}), which can artificially increase the fraction of irregular galaxies. However, since HST and JWST-based classifications are shown to agree reasonably well, we might deduce that the impact of this bias is not dominant.

\section{Summary and conclusions}
\label{sec:summary}
In this work, we present a morphological classification of around 20,000 galaxies with $F200W[AB]<27$ using JWST/NIRCam images in four different filters - $F150W$, $F200W$, $F356W$, and $F444W$ - obtained from the CEERS survey. We classify galaxies into four classes - spheroid, bulge+disk, disk, and irregular - using a Convolutional Neural Network (CNN) and Adversarial Domain Adaptation. The resulting classification shows excellent agreement with independent visual classifications, demonstrating the successful adaptation of the CNN to the new domain and will be made publicly available as part of the CEERS data products.

We compare our JWST-based classifications with existing HST/WFC3-based classifications. We find that $\sim90\%$ and $\sim75\%$ of galaxies with $z<3$ have the same early/late and regular/irregular class respectively in both JWST and HST imaging when considering similar wavelengths. For the smallest and faintest objects, NIRCam based classifications tend to find fewer bulge-dominated and disturbed galaxies, likely due to a combination of SNR and spatial resolution. However, the impact on the measured morphological fractions as a function of cosmic time is minimal.

In the second part of the study, we analyze the rest-frame NIR ($\sim 0.8 - 1\mu m$) morphologies of a mass-complete sample ($\log M_*/M_\odot>9$) of galaxies from $z\sim6$ to $z\sim0$. Our findings include:

\begin{itemize}
\item The fraction of bulge-dominated galaxies increases at the high-mass end, even at $z\sim5$, indicating that the processes of bulge formation in massive galaxies are already in place at these early cosmic epochs.
\item The fraction of peculiar galaxies also increases with redshift, even in the NIR rest-frame, suggesting that the stellar mass distribution is more disturbed at high redshift, although the SNR may still affect this result.
\item The high-mass end of the galaxy distribution ($\log M_*/M_\odot>10.5$) is dominated by undisturbed disk-like morphologies even at $z\sim5$, indicating that disk formation may be in place at very early epochs.
\item The fraction of early-type galaxies reaches $\sim70\%$ to $\sim90\%$ for massive ($\log M_*/M_\odot>10.5$) quenched galaxies, even at $z\sim5$, suggesting that the connection between quenching and bulge growth is already established around $\sim1$ Gyr after the Big Bang.
\end{itemize}

Overall, our results indicate a complex morphological diversity already in place $\sim 1$ Gyr after the Big-Bang.

\begin{acknowledgements} 
MHC, EA, RS and JVF acknowledge financial support from the State Research Agency (AEI\-MCINN) of the Spanish Ministry of Science and Innovation under the grants ``Galaxy Evolution with Artificial Intelligence" with reference PGC2018-100852-A-I00 and "BASALT" with reference PID2021-126838NB-I00. 
\end{acknowledgements}

\bibliographystyle{aa}
\bibliography{biblio.bib}

\begin{appendix}
\section{Star-Galaxy separation}
\label{app:star-galaxy}

\begin{figure}[h]
\includegraphics[width=1\linewidth]{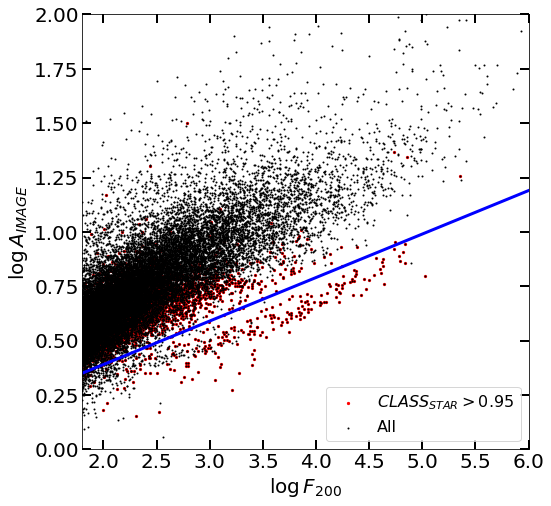}
    \caption{$\log F_{200}-\log A_{IMAGE}$ plane used in this work for star-galaxy separation. The black dots show all the sample of selected objects with F200W AB apparent magnitude smaller than 27. The red points are those objects with a value of $CLASS_{STAR}>0.95$. The blue solid line show the boundary used to separate stars (below the line) from galaxies (above the line).} 
    \label{fig:star_galaxy}
\end{figure}

We use a simple procedure to perform a rough star-galaxy separation in the F200W filter using three parameters delivered by \textsc{SExtractor}: $A_{IMAGE}$, $F_{200}$ and $CLASS_{STAR}$ which measure the isophotal image major axis, the isophotal flux and a stellarity flag computed with a pre-trained Neural Network, respectively. In Figure~\ref{fig:star_galaxy} we plot $\log F_{200}$ vs. $\log A_{IMAGE}$ for all galaxies in our sample and for those with a value of $CLASS_{STAR}$ larger than 0.95. We see that the latter tend to be located in a well defined sequence separated from the bulk of the distribution. We thus identify this sequence as the locus of bright stars and draw an empirical linear relation below which objects are flagged as stars. While this recipe might work reasonably well to identify bright stars, the separation is expected to be much less obvious at faint fluxes as seen in Figure~\ref{fig:star_galaxy}. However, given that the main analysis of this work focuses on galaxies more massive than $10^9$ solar masses, we do not expect the contamination of stars to significantly affect any of the conclusions.

\section{HST vs. JWST stamps}
\label{sec:hst_jwst}
\begin{figure*}
    \begin{minipage}[t]{0.5\linewidth}
        \centering
        \captionsetup{font=footnotesize}
        \caption*{CEERS late-type [F150W]}
        \vspace{-11pt}
        \includegraphics[width=\linewidth]{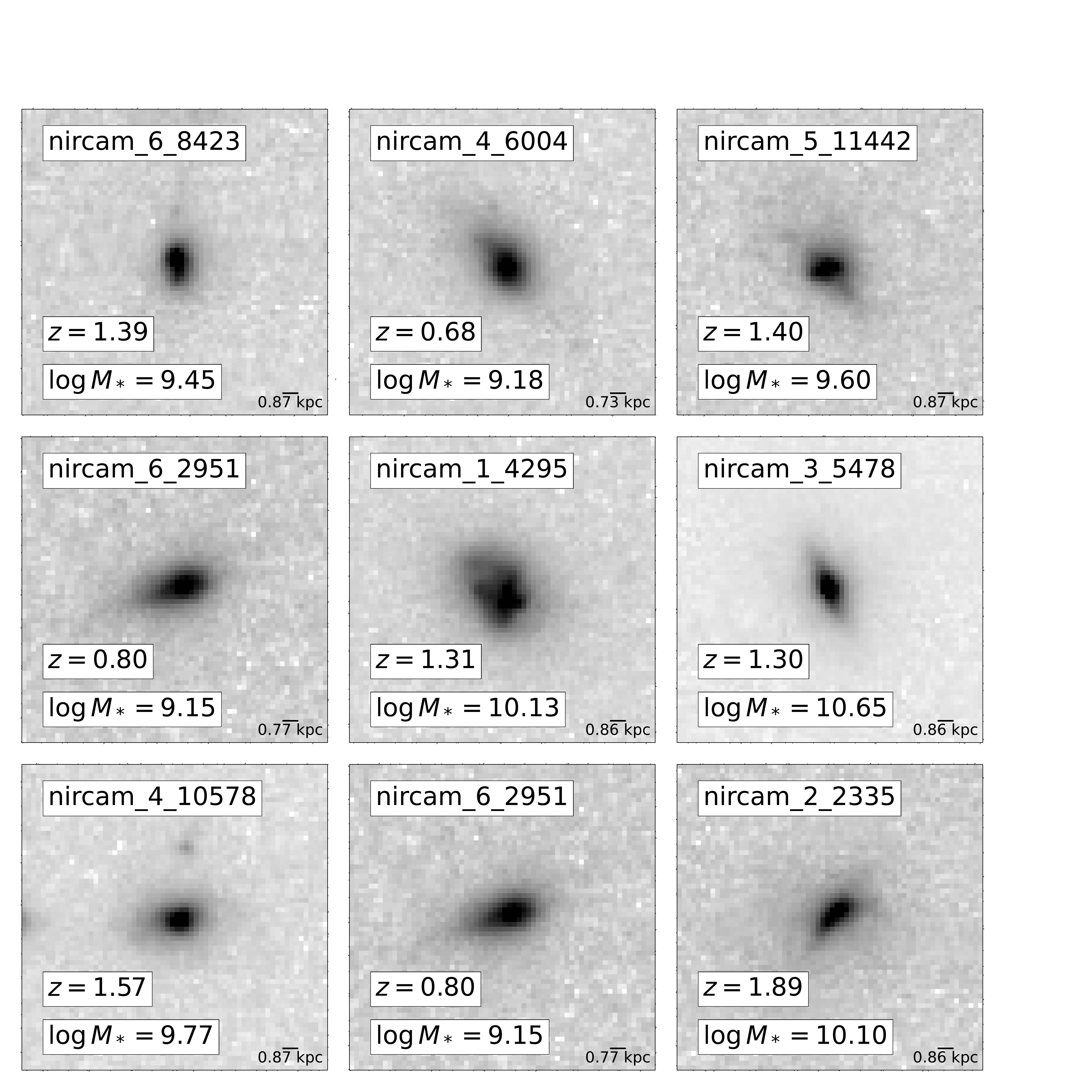}
    \end{minipage}%
    \begin{minipage}[t]{0.5\linewidth}
        \centering
        \captionsetup{font=footnotesize}
        \caption*{CANDELS early-type [F160W]}
        \vspace{-11pt}
        \includegraphics[width=\linewidth]{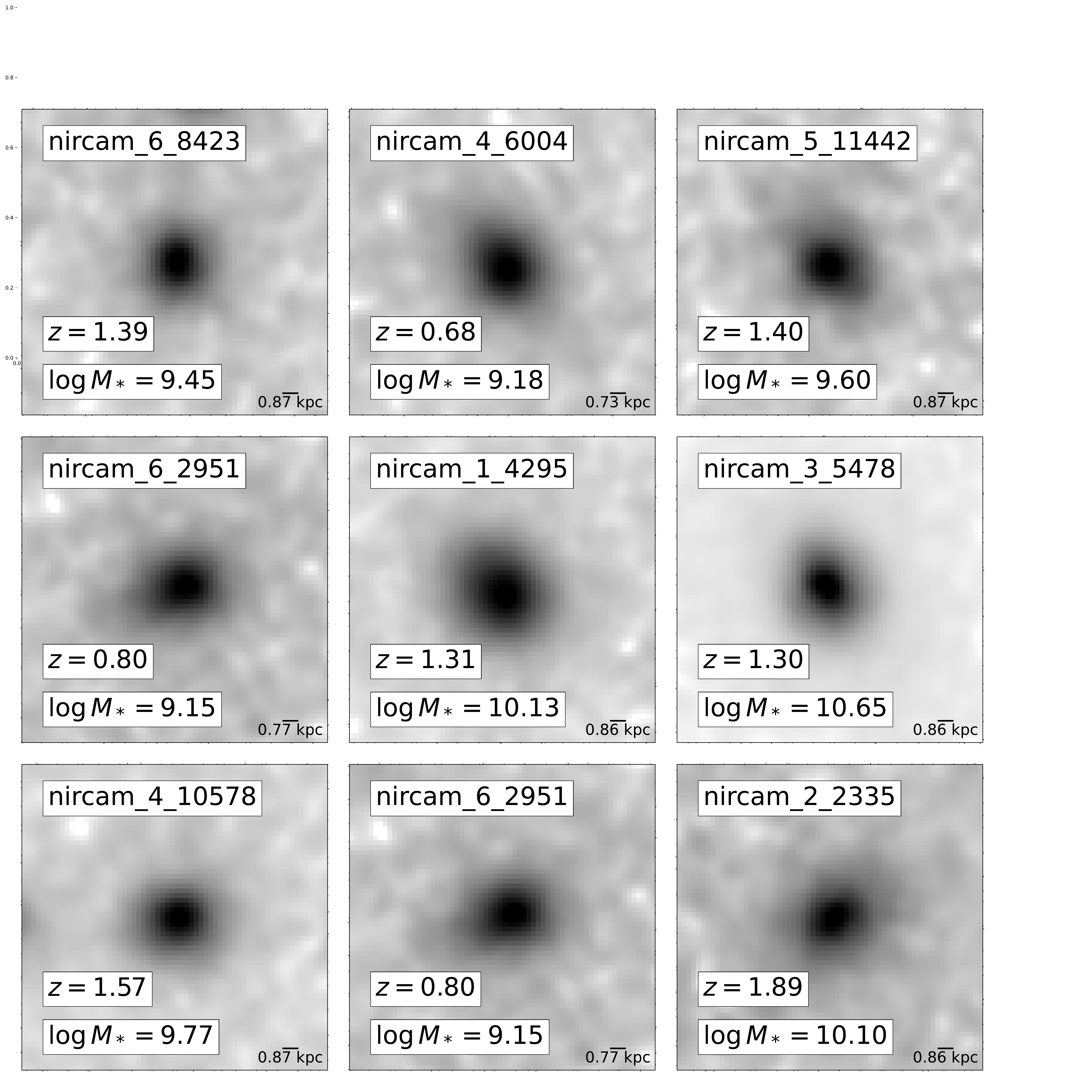}
    \end{minipage}
    
    \begin{minipage}[t]{0.5\linewidth}
        \centering
        \captionsetup{font=footnotesize}
        \caption*{CEERS disks [F150W]}
        \vspace{-11pt}
        \includegraphics[width=\linewidth]{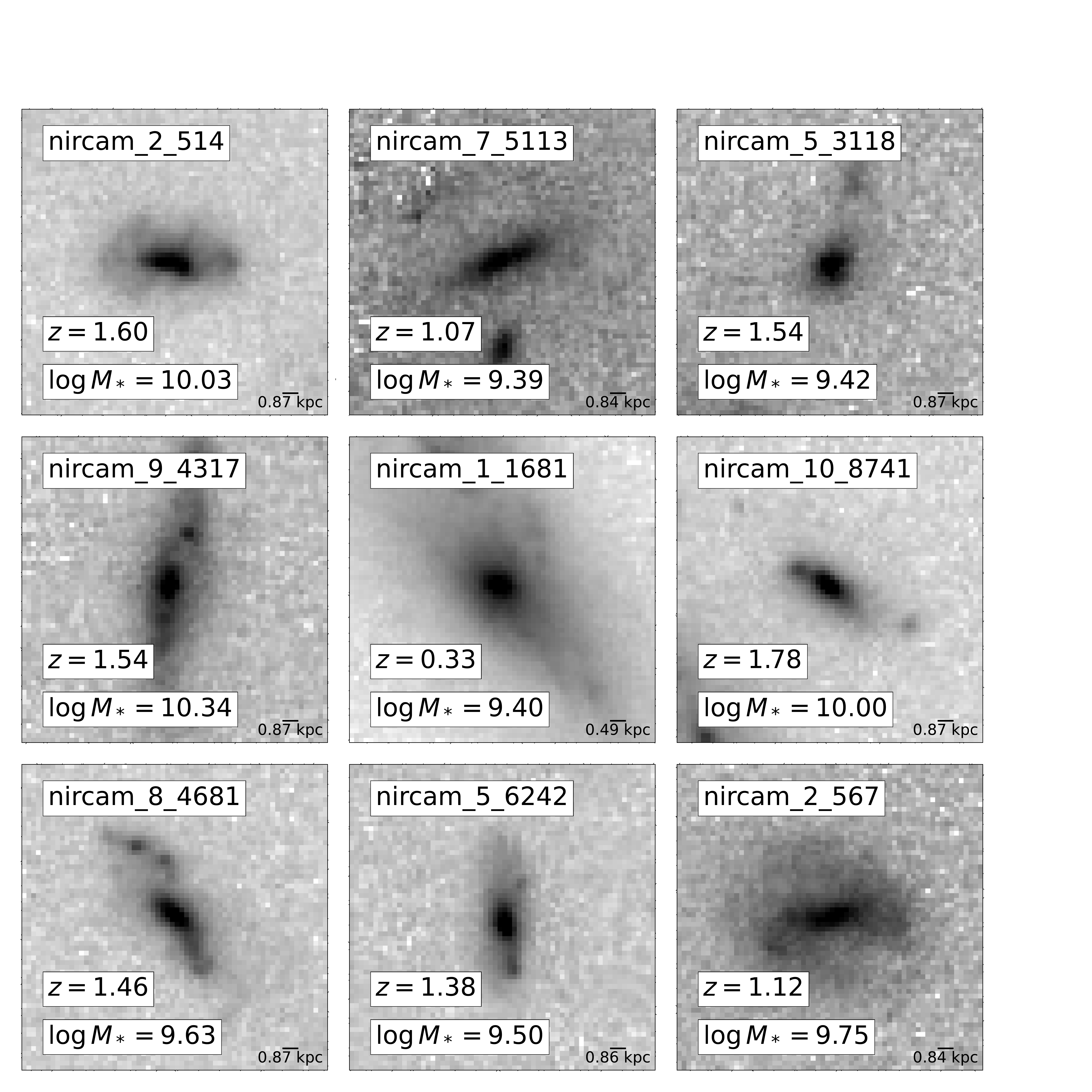}
    \end{minipage}%
    \begin{minipage}[t]{0.5\linewidth}
        \centering
        \captionsetup{font=footnotesize}
        \caption*{CANDELS irregular [F160W]}
        \vspace{-11pt}
        \includegraphics[width=\linewidth]{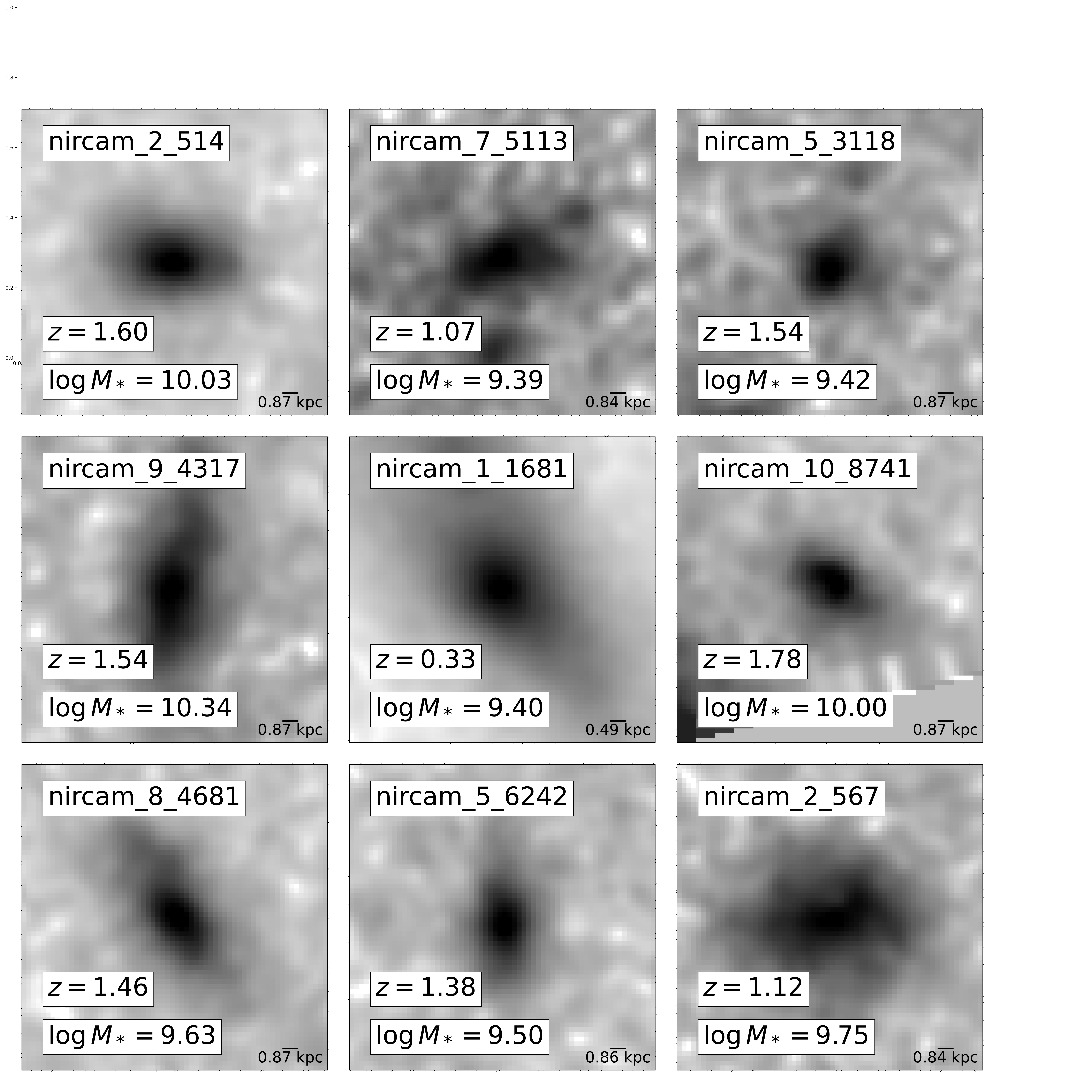}
    \end{minipage}
    
    \captionsetup{font=large}
    \caption{Example of random stamps of the same galaxies in CEERS (F150W) and CANDELS (F160W) at the same pixel scale with different morphological classifications. The two top panels show galaxies classified as late-type in CEERS (top left) and as early-type in CANDELS (top right). The bottom panels show galaxies classifed as disks in CEERS (botttom left) and irregular in CANDELS (bottom right). The physical scale in kpc in shown for every galaxy. A square root scaling has been applied to enhance the outskirts.} 
    \label{fig:CEERS_CANDELS_sph}
\end{figure*}

We show in Figure~\ref{fig:CEERS_CANDELS_sph} some examples of galaxies with different morphological classifications in CEERS and CANDELS observed at a similar wavelength and at the same pixel scale of $0.03^{"}/pixel$. We can see how resolution and depth can impact the classification. Objects which change the classification for early-type in CANDELS to late-type in CEERS are typically small as shown in Figure~\ref{fig:hex_CEERS_CANDELS} and therefore appear rounder in CANDELS. Regarding objects which are classified as irregular in CANDELS but not in CEERS, we see there are more extended and the better sensitivity and resolution of CEERS enables a clearer detection of a diffuse component which can be interpreted as a disk. Nevertheless, galaxies still tend to present some asymmetries which illustrates the difficulty of defining the irregular class.  

\section{Spatial resolution and rest-frame wavelength}

\begin{figure}
    {\includegraphics[width=\linewidth]{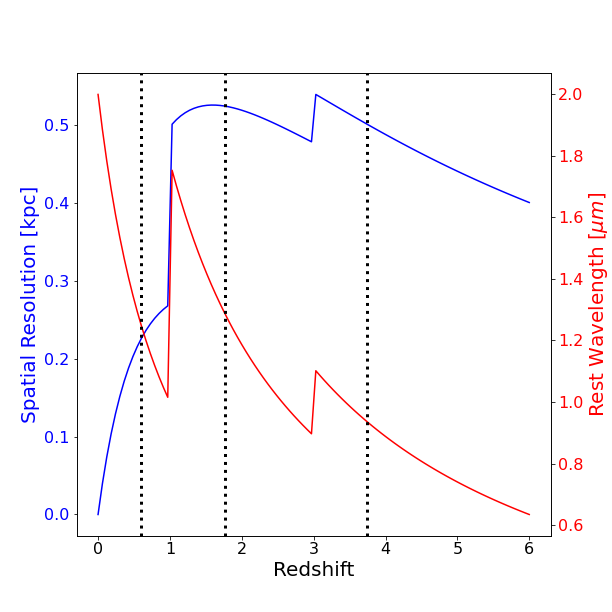}}
      \caption{Spatial resolution in kpc (blue solid line) and rest wavelength (red solid line) as a function of redshift for the three different filters used in this work: $F200W$ at $z<1$, $F356W$ at $1<z<3$ and $F444W$ at $z>3$. The dashed vertical lines indicate the mean redshift in each of three redshift bins used throughout this work. } 
    \label{fig:rest_wl}
\end{figure}

In this work we try to quantify the evolution of galaxy morphology in the same rest-frame wavelength by using different filters depending on the redshift bin.  This implies that the effective spatial resolution also changes. In figure~\ref{fig:rest_wl} we show the evolution of the rest-frame wavelength and spatial resolution for the three different filters used in this work. The spatial resolution is computed as the Full Width Half Maximum (FWHM) value for every filter reported in the Table 1 of~\cite{2023ApJ...946L..13F} and divided by $2.35$.

\end{appendix}

\end{document}